\documentclass{elsart_TR2}

\usepackage{latexsym}
\usepackage{graphicx,amssymb,psfrag}
\usepackage{amsmath}
\usepackage{enumerate}
\usepackage{varioref}
\usepackage{rotating}
\usepackage{graphics}
\usepackage[dvips]{epsfig}
\usepackage{amssymb}
\usepackage{url}

\def\boxit#1{\vbox{\hrule\hbox{\vrule\kern4pt
  \vbox{\kern1pt#1\kern1pt}
\kern2pt\vrule}\hrule}}

\newcommand\nc{\newcommand}

\newtheorem{theorem}{\bfseries Theorem}
\newtheorem{lemma}{\bfseries Lemma}
\newtheorem{rull}{\bfseries Rule}
\newtheorem{corollary}{\bfseries Corollary}
\newtheorem{definition}{\bfseries Definition}

\nc{\crl}[2]{\begin{corollary}\label{crl:#1} #2 \end{corollary}}
\nc{\dfn}[2]{\begin{definition}\label{def:#1} #2 \end{definition}}
\nc{\llem}[2]{\begin{lemma}\label{lem:#1} #2 \end{lemma}}
\nc{\thmm}[2]{\begin{theorem}\label{thm:#1} #2\end{theorem}}
\nc{\rul}[2]{\begin{rull}\label{rull:#1} #2\end{rull}}

\nc{\eqn}[2]{\begin{eqnarray}\label{eqn:#1} #2 \end{eqnarray}}

\nc{\fig}[4]{\begin{figure}[h]
\begin{center}
\includegraphics[width=#2\textwidth]{#4}
\end{center}
\caption{#3}\label{fig:#1}
\end{figure}}

\nc{\tbl}[3]{\begin{table}[hbt] #3 \caption{#2} \label{tab:#1}
\end{table}}

\nc{\refc}[1]{Corollary~\ref{crl:#1}}
\nc{\refd}[1]{Definition~\ref{def:#1}}
\nc{\reff}[1]{Figure~\ref{fig:#1}}
\nc{\refl}[1]{Lemma~\ref{lem:#1}}
\nc{\refp}[1]{Proposition~\ref{prp:#1}}
\nc{\reft}[1]{Theorem~\ref{thm:#1}} \nc{\refe}[1]{(\ref{eqn:#1})}
\nc{\reftb}[1]{Table~\ref{tab:#1}}
\nc{\refr}[1]{Rule~\ref{rull:#1}}

\nc{\reffc}[1]{Fact~\ref{fact:#1}}

\nc{\pff}[1]{ \noindent \emph{Proof.} #1 \hfill \qed\par}

\long\def\invis#1{}

\begin{document}
\begin{frontmatter}
\title{Exact Algorithms for Dominating Induced Matching Based on Graph Partition
}
\author{Mingyu Xiao\corauthref{cor}}
\corauth[cor]{Corresponding author. Tel.: +86 15397626165}
\ead{myxiao@gmail.com}
\address{School of Computer Science and Engineering\\
University of Electronic Science and Technology of China\\
Chengdu 611731, China}
\author{Hiroshi Nagamochi}
 \ead{nag@amp.i.kyoto-u.ac.jp}
\address{Department of Applied Mathematics and Physics\\
Graduate School of Informatics\\
Kyoto University\\
Yoshida Honmachi, Sakyo, Kyoto 606-8501, Japan}

\begin{abstract}
A \emph{dominating induced matching}, also called an \emph{efficient edge domination}, of a graph $G=(V,E)$ with $n=|V|$ vertices and $m=|E|$ edges
is a subset $F \subseteq E$ of edges in the graph such
that no two edges in $F$ share a common endpoint and each edge in $E\setminus F$ is incident with exactly one edge in $F$. It is NP-hard to decide whether a graph admits a dominating induced matching or not. In this paper, we design a $1.1467^nn^{O(1)}$-time exact algorithm for this problem, improving all previous results.
This problem can be redefined as a partition problem that is to partition the vertex set of a graph into two parts $I$ and $F$, where $I$ induces an independent set (a 0-regular graph) and $F$ induces
 a perfect matching (a 1-regular graph).
After giving several structural properties of the problem, we show that the problem always contains some ``good vertices,'' branching on which by including them to either $I$ or $F$ we can effectively reduce the graph. This leads to a fast exact algorithm to this problem.

\end{abstract}

\end{frontmatter}

\section{Introduction}

Let $G=(V,E)$ be a simple undirected graph. An edge $e\in E$ \emph{dominates} itself and all edges sharing a common endpoint with it.
A subset $F\subseteq E$ of edges is called an \emph{edge domination} if each edge in the graph is dominated by at least one edge in $F$.
An edge domination $F$ is called \emph{efficient} if each edge in the graph is dominated by exactly one edge in $F$.
A graph has many edge dominations but may not have an efficient edge domination.
To  find an optimal edge domination or an efficient edge domination in a graph is a hard problem and this problem finds
applications in the fields of design and analysis of communication networks, network routing and coding theory~\cite{CCDS:EEDRegularGraph,GSSH:EED}.

Note that an efficient edge domination is an induced matching in the graph such that each  edge is dominated by exactly one edge in it.
So an efficient edge domination is also called a \emph{dominating induced matching}. The dominating induced matching problem (\textsc{Dominating Induced Matching}),
also called \textsc{Efficient Edge Domination} in the literature, is to check whether a graph admits a dominating induced matching or not. This problem has been extensively studied, especially
in terms of the computational complexity of it in different graph
 classes~\cite{BHN:EEDhole-free,BM:DIMp7,CKL:DIM,GSSH:EED,PandEED,EED:bpermutation}.
Grinstead et al.~\cite{GSSH:EED} first established the NP-hardness of this problem.
Later it is was further shown to be NP-hard even in planar bipartite graphs of maximum degree 3~\cite{BHN:EEDhole-free} and $d$-regular graphs for $d\geq3$~\cite{CCDS:EEDRegularGraph}.
On the other hand, this problem is polynomial-time solvable in  many graph classes such as
AT-free graphs~\cite{BLR:EEDhypergraph}, hole-free graphs~\cite{BHN:EEDhole-free}, $P_7$-free graphs~\cite{BM:DIMp7},
and claw-free graphs~\cite{CKL:DIM}.

In terms of exact algorithms, the edge domination problem,
a problem of finding an edge domination of minimum size, has been extensively studied~\cite{kn:rooij,xn:eds}. Most exact algorithms for the edge domination problem are analyzed by using the measure-and-conquer method. However, none of them can be easily modified for
\textsc{Dominating Induced Matching}.
Some exact algorithms for \textsc{Dominating Induced Matching} were also introduced recently~\cite{LMS:DIMexact,LMS:DIMexact_odd}. Lin et al.~\cite{LMS:DIMexact} obtained a $1.1939^nn^{O(1)}$-time algorithm. Their algorithm branches on a vertex by including it to the vertex set of the dominating induced matching or not.
In this paper, we also use this idea to design a branch-and-reduce algorithm. However, our improvement is not obtained by carefully
checking the worst cases of previous algorithms. We derive several graph properties of the problem, which will reduce some bad cases and allow us to design an improved algorithm.

Our paper is organized as follows. The notation system and our algorithm type are introduced in Section~\ref{sec_pre}.
Some conditions for feasibility and several rules to simplify problem  instances are given in Section~\ref{sec_rules}. The full algorithm is described in Section~\ref{sec_alg} and
the detailed analysis is given in Section~\ref{sec_analy}.
Finally, some concluding remarks are put in Section~\ref{sec_con}.

\section{Preliminaries}
\label{sec_pre}
In this paper, a graph stands for a simple undirected graph.
Let $G=(V,E)$ be a graph and $X\subseteq V$ be a subset of vertices.
A vertex in $X$ is called an \emph{$X$-vertex},
and a neighbor $u\in X$ of a vertex $v$ is called an \emph{$X$-neighbor} of $v$.
The subgraph induced by $X$ is denoted by $G[X]$, and $G[V\setminus X]$ is also written as $G\setminus X$.
For a subset $E'\subseteq E$, let $G-E'$ denote the subgraph obtained from $G$ by deleting edges in $E'$.
The vertex set and edge set of a graph $G$ are denoted by $V(G)$ and $E(G)$, respectively.
A vertex $v$ (resp., an edge $e$) in a graph is called a \emph{cut-vertex} (resp., \emph{bridge})
 if removing it increases the number of connected components of the graph.
For a vertex $v\in V$,  the set of vertices with distance $k$ from $v$ is denoted by $N_k(v)$, where $N_0(v)=\{v\}$ and $N_1(v)$ is also simply written as $N(v)$. The \emph{degree} of vertex $v$ is defined to be $|N(v)|$.
For an edge $uv \in E$, let $N_0(uv)=\{u,v\}$,
 and $N_k(ab)=N_k(a)\cup N_k(b)\setminus\cup_{i=0}^{k-1}(N_i(a)\cup N_i(b))$ for $k\geq 1$,
where $N_1(uv)=N(u)\cup N(v)\setminus \{u,v\}$ is also written as $N(uv)$.
For a vertex $x=v$ or an edge $x=ab$ and an integer $i\geq 1$, we let $N_i[x]=\cup_{j=0}^i N_j(x) $.
A path (resp., cycle) of \emph{length} $k$, i.e., a path (resp., cycle) containing exactly $k$ edges is called
 a  \emph{$k$-path} (resp., \emph{$k$-cycle}).
When a given graph is  edge-weighted, we use $w(e)$ to denote the weight of edge $e$.
For a graph with no edges, we treat the empty edge subset as a dominating induced matching.

Some references, such as~\cite{LMS:DIMexact},  address the weighted case of \textsc{Dominating Induced Matching}, in which the edges in the graph have a cost and the goal of the problem is to find
a dominating induced matching of minimum cost among all dominating induced matchings (if they exist). To find a dominating induced matching of minimum cost or maximum cost is NP-hard, since it is NP-hard to check the existence of dominating induced matchings in an unweighted graph. We call these two weighted versions of this problem the \emph{minimum version} and the \emph{maximum version}. Many search algorithms for the unweighted version, including our algorithms, can be easily modified to the weighted versions. So in this paper, we describe our algorithms in terms of the unweighted version and point out the arguments where the weighted versions may need additional operations for handling edge weights.
In our algorithm, non-negativeness of edge weights $w$ is not necessary to be assumed.
We treat only the minimum version because  the maximum version can be solved
as the minimum version  just by replacing $w$ with $-w$.

\subsection{Algorithms based on graph partition}
\textsc{Dominating Induced Matching} can also be defined as a partition problem: Whether the vertex set of a graph can be partitioned into two subsets $I$ and $F$ such that
 $I$ induces an independent set (a graph with  degree-0 vertices) and
 $F$ induces a matching (a graph with   degree-1 vertices).

We can solve  \textsc{Dominating Induced Matching} in $2^nn^{O(1)}$ time by checking all partitions of the vertex set.
By using some branch-and-reduce methods, the searching space can be reduced greatly~\cite{LMS:DIMexact}.
Our algorithm is also a branch-and-reduce algorithm. We fix some vertices in the two sides of the partition and then try to extend them by some effective operations.

\medskip
For a subset $M\subseteq V$ and an independent set
$I\subseteq V\setminus M$ in $G$,
a dominating induced matching $F$ is called an {\em $(M,I)$-dim}
if
\[M\subseteq V(F) \subseteq V\setminus I.\]
We may use $(G,M,I)$ to denote an instance of the problem to decide whether the graph $G$ admits an $(M,I)$-dim or not.
We always let $U$ denote $V\setminus (M\cup I)$,
and let $M_i$ denote the set of $M$-vertices $u$ such that $|N(u)\cap M|=i$.
The vertices in $U$ are called {\em undecided\/} vertices.
After setting $M=I=\emptyset$ and $U=V$ initially,
we search for an $(M,I)$-dim $F$,
 keeping track of subsets $M$ and $I$ of the vertices of $G$.
The following \emph{Basic Conditions} are kept invariant.
\begin{enumerate}
\item
$I$ is an independent set in $G$;
\item
$M=M_0\cup M_1$, i.e., no $M$-vertex has two or more $M$-neighbors; and
\item
Each $M_0$-vertex has  at least one $U$-neighbor.
\end{enumerate}
We specify an instance only by $G$, $M$ and $I$, from which
$U$, $M_0$ and $M_1$ are uniquely determined.

When there are no undecided vertices in the graph, i.e., $U=\emptyset$,
we can easily know whether or not the  current graph has an $(M,I)$-dim $F$ by checking if it satisfies the Basic Conditions.
In our algorithm, we use reduction and branching rules to move
$U$-vertices to either $M$ or $I$ until $U$ becomes an empty set. In a reduction rule, we move some $U$-vertices  to $M\cup I$ directly keeping the optimality of the solution.
In a branching rule, we generate two subinstances by moving a $U$-vertex to either $M$ or $I$. A branch-and-reduce algorithm consisting of reduction and branching rules will generate a search tree.
Each node of the search tree represents an instance in the algorithm.
In particular, the instance to the root of the search tree is the initial instance, and
 the instances to the leaves of the search tree are instances with $U=\emptyset$.

\subsection{Branch-and-reduce algorithms and recurrence relations}

Our algorithm contains one branching rule that is to branch by moving a $U$-vertex to either $M$ or $I$.
After  moving a $U$-vertex to $M$ or $I$, we may be able
to move some other $U$-vertices to $M$ or $I$ directly by some reduction rules.
We assume that the number of $U$-vertices decreases by at least $a$ and $b$ in the two resulting instances, respectively. Let $C(n)$ denote the number of the leaves of the search tree generated by the algorithm to solve a problem with $n$ $U$-vertices. Then we get the following recurrence relation:
$$C(n)\leq C(n-a)+C(n-b).$$
To derive an upper bound on the size of the search tree, or the exponential part of the running time bound of the algorithm, we need to solve
this kind of linear recurrence relation.
A solution to the above recurrence relation is of the form $C(n)=O(\alpha^n)$, where $\alpha$ is the unique positive real root of the function $x^n-x^{n-a}-x^{n-b}=0$.
We call $\alpha$ the \emph{branching factor} of $C(n)\leq C(n-a)+C(n-b)$.
For the largest branching factor  $\alpha$ in the algorithm,
 the size of the search tree is $O(\alpha ^n)$.
For more details about how to  evaluate the size of the search tree and to solve the recurrence relations,
the readers are referred to the monograph~\cite{Fomin:book}.


We introduce a notation to describe a relationship among  recurrence relations, which will be used for us to ignore some recurrence relations without missing the worst recurrence relations with the largest branching factor.
 For two  recurrence relations
$A:~C(n)\leq C(n-a_1)+C(n-a_2)+\cdots+ C(n-a_t)$ and
$B:~C(n)\leq C(n-b_1)+C(n-b_2)+\cdots+ C(n-b_t)$ with the same number $t$ of branches,
we denote $A\leq_c B$,
if
there are indices $i_1,i_2\in \{1,2,\ldots,t\}$ such that $a_{i_1}+a_{i_2}\geq b_{i_1}+b_{i_2}$, $b_{i_1}\geq a_{i_1}\geq a_{i_2} \geq b_{i_2}$ and $a_i=b_i$ for
 all $i\in \{1,2,\ldots,t\}\setminus \{i_1,i_2\}$.
The recurrence relation $A$ is \emph{covered} by the recurrence relation $B$ if $A\leq_c B$ or there is a finite sequence of recurrence relations
$A_1,A_2,\ldots, A_l$ such that
$A \leq_c A_1 \leq_c A_2 \leq_c \cdots \leq_c A_l \leq_c B$.
We see that the branching factor of a recurrence relation is not smaller than that of any recurrence relation covered by it~\cite{Fomin:book}.

We may also derive a single recurrence relation for a series of branching operations
by combining the recurrence relations of the operations.
Given two recurrence relations
$X:~C(n)\leq C(n-x_1)+C(n-x_2)+\cdots+ C(n-x_t)$ and
$Y:~C(n)\leq C(n-y_1)+C(n-y_2)+\cdots+ C(n-y_s)$,
let  $X_Y$ denote
the combined recurrence relation of the branching with $X$ and then
branching with $Y$ in the first branch in $X$;
 i.e., $X_Y$ is $C(n)\leq \sum_{1\leq i\leq s}C(n-x_1-y_i)+C(n-x_2)+\cdots+ C(n-x_t)$.
The following lemma allows us to ignore some recurrence relations covered
by others in our algorithm to find the largest branching factor.

\llem{combined_recurrences}{Given three recurrence relations
$A:~C(n)\leq C(n-a_1)+C(n-a_2)+\cdots+ C(n-a_t)$,
$B:~C(n)\leq C(n-b_1)+C(n-b_2)+\cdots+ C(n-b_t)$ and
$D:~C(n)\leq C(n-d_1)+C(n-d_2)+\cdots+ C(n-d_s)$.
Assume that $A$ is covered by $B$. Then the combined recurrence relation $D_A$ is covered by the combined recurrence relation $D_B$.
}

\pff{
If $A\leq_c B$ then $D_A \leq_c D_B$, because
for the indices
$i_1,i_2\in \{1,2,\ldots,t\}$ such that $a_{i_1}+a_{i_2}\geq b_{i_1}+b_{i_2}$, $b_{i_1}\geq a_{i_1}\geq a_{i_2} \geq b_{i_2}$ and $a_i=b_i$ for  all $i\in \{1,2,\ldots,t\}\setminus \{i_1,i_2\}$,
we see that
$(d_1+a_{i_1})+(d_1+a_{i_2})\geq (d_1+b_{i_1})+(d_1+b_{i_2})$, $d_1+b_{i_1}\geq d_1+a_{i_1}\geq d_1+a_{i_2} \geq d_1+b_{i_2}$ and $d_1+a_i=d_1+b_i$
for all
$i\in \{1,2,\ldots,t\}\setminus \{i_1,i_2\}$.
Otherwise, there is a sequence of recurrence relations $A_1,A_2,\ldots, A_l$ such that
 $A_0=A$, $A_{l+1}=B$, and
$A_i \leq_c A_{i+1}$ for each
$i=0,1,\ldots,l$.
Analogously with the case of $A\leq_c B$,
 we have $D_A \leq_c D_{A_1} \leq_c D_{A_2} \leq_c \cdots \leq_c D_{A_l} \leq_c D_B$.
 }\medskip

In our algorithm, $C(n)\leq C(n-2)+C(n-8)$ is one of the worst recurrence relations, which solves to $C(n)=O(1.1749^n)$.
However, we observe that the worst cases will not always happen in our algorithm. After branching with a worst recurrence,
we can branch with a much better recurrence in the next step. So we combine the bad and good recurrences together to get a single recurrence.
Finally we get an upper bound $O(1.1467^n)$ on the size of the search tree.

In the next section, we first introduce some properties of \textsc{Dominating Induced Matching},
which show that some $U$-vertices can be moved to $M\cup I$ directly and will be used to design reduction rules for our algorithm.
We design our algorithm so that it returns an  $(M,I)$-dim of a given instance $(G,M,I)$ if any or
0 as a message of the infeasibility otherwise.

\section{Properties and Reduction Rules}
\label{sec_rules}

In this section, we will give some rules to reduce a given instance.
A reduction rule is called \emph{correct} if it preserves the feasibility of instances; i.e.,
for an instance $\mathcal{I}=(G,M,I)$ and the instance  $\mathcal{I}'=(G',M',I')$ obtained from $\mathcal{I}$ by applying the reduction rule,
   $\mathcal{I}$ is an yes-instance if and only if so is $\mathcal{I}'$. We will show the correctness of our reduction rules.

Clearly every instance $(G,M,I)$ violating the Basic Conditions cannot have an $(M,I)$-dim.
This provides the following reduction rule.

\rul{rule0}{When the current instance violates the Basic Conditions, halt and return 0 to indicate that there is no $(M,I)$-dim.}

In the search steps of our algorithm, we can ignore any resulting instance that violates
the Basic Conditions. We also execute the following reduction operations to move $U$-vertices  to $M$ or $I$ without branching.
\rul{rule1}{Move to $I$ a $U$-vertex $v$ adjacent to some $M_1$-vertex.} 
\rul{rule2}{Move to $M$ a $U$-vertex $v$ adjacent to some $I$-vertex.} 
\rul{rule3}{Move to $M$ the unique $U$-neighbor $v$ of some $M_0$-vertex $u$, i.e.,  $\{v\}=N(u)\cap U$.}

A $U$-vertex $v$ in an instance $(G,M,I)$ is called \emph{i-reducible} (resp., \emph{m-reducible})
if  moving it to $M$ (resp., $I$) and applying Rules~2 to 4 as much as possible result in an instance that violates the Basic
Conditions.
Every i-reducible vertex should not be in the vertex set of any $(M,I)$-dim of the instance $(G,M,I)$,
whereas every m-reducible vertex should be in the vertex set of any $(M,I)$-dim of the instance $(G,M,I)$ if it exists.
A vertex is called \emph{infeasible} if it is both i-reducible and m-reducible.
Clearly every instance with some infeasible vertex admits no $(M,I)$-dim.
Whether a vertex is i-reducible (resp., m-reducible) or not can be checked in polynomial time. We have the following rules:

\rul{rule6}{When there is an infeasible vertex, halt and return 0 to indicate that there is no $(M,I)$-dim.}
\rul{rule4}{Move any i-reducible vertex to $I$.}
\rul{rule5}{Move any m-reducible vertex to $M$.}

Next we identify some infeasible,   i-reducible and  m-reducible  vertices from graph structures.

Obviously any dominating induced matching contains exactly one vertex in a triangle,
whereas no dominating induced matching contains any edge in a 4-cycle.
Hence a complete graph with size 4 cannot admit a dominating induced matching.
\llem{k4}{\emph{\cite{LMS:DIMexact}} A $U$-vertex in a set of four vertices that induces a clique of size $4$   is infeasible.}

\llem{i-reducible}{A $U$-vertex $v$  is i-reducible if \\
{\rm (a)}  $v$ has at least two $M$-neighbors; or \\
{\rm (b)} $G[U]$ contains two triangles $v a a'$ and $v b b'$ with $\{a,a'\}\cap \{b,b'\}=\emptyset$.}

\llem{i-reducible2}{ A $U$-vertex  $v$ adjacent to an $M_0$-vertex $u$
is i-reducible if there are two adjacent  $U$-vertices $v_1,v_2\in N(v)\cup N(u)\setminus\{v\}$; i.e.,
$u v_1 v_2$ or
 $v v_1 v_2$  is a triangle,  or
$v u v_1 v_2$ is a 4-cycle.}

\llem{m-reducible}{A $U$-vertex $v$ is m-reducible if \\
{\rm (a)} $v$ is a unique $U\cup M$-neighbor of some $U$-vertex $u$, i.e., $\{v\}=N(u)\cap (U\cup M)$; or  \\
{\rm (b)} $G$ has a 4-cycle $v u v' u'$ such that  $vv'$ is a chord or $u$ is of degree 2.
}

We call an instance  {\em pseudo-feasible} if none of the above  seven rules is applicable.
When applying any other kind of reduction and branching rules, we always assume that the current instance is pseudo-feasible.

\medskip
It is easy to observe the next.
\llem{pseudo}{Let $(G,M,I)$ be a pseudo-feasible instance 
and  $(X_1,X_2,X_3,X_4)$ be the sequence $(U,M_0,I,M_1)$ of vertex subsets in the instance.
Then:\\
{\rm (i)} There is no edge  between $X_i$ and $X_{j}$ with $|i-j|\geq 2$; and \\
{\rm (ii)} $(G,M,I)$ is an yes-instance if and only if
for each connected component $H$ of $G[U\cup M_0]$ the instance $(H, M_0\cap V(H), \emptyset)$ is an yes-instance.}
%
\crl{cut}{In a pseudo-feasible instance, if the graph $G$ is connected and $M$ is not an empty set, then each connected component of $G[U\cup M_0]$ contains at least one $M_0$-vertex.}

Note that when we regard $(H, M_0\cap V(H), \emptyset)$ as a new instance $(G,M,I)$,
the associated vertex sets $M_1$ and $I$ are empty.
Then the lemma provides a method of decomposing an instance into those with connected graphs
and no vertices in $I\cup M_1$.
From now on, we consider a  pseudo-feasible instance $(G,M,I)$ with $I=M_1=\emptyset$ such that
$G$ is connected.

%


\llem{remove_edge}{Let $v$ and $v'$ be two vertices in an  instance $(G,M,I)$  such that
any $(M,I)$-dim $F$ in $G$ satisfies $|V(F)\cap \{v, v' \}|=1$.
Let $G'$ be the graph obtained by deleting the edge $vv'$ from $G$ if $vv'\in E(G)$
 or by adding an edge $vv'$ to $G$ if  $vv'\not\in E(G)$.
If $(G,M,I)$ is an yes-instance, then $(G',M,I)$ is also an yes-instance.
}
\pff{Since $v$ and $v'$ belong to $V(F)$ and $V(G)\setminus V(F)$ separately in any $(M,I)$-dim $F$ of $G$,
the edge set $F$ is also an $(M,I)$-dim of $G'$. }\medskip

\llem{tri1}{Let   $(G,M,I)$ be a pseudo-feasible  instance with $I=M_1=\emptyset$,
and let $u$ be a degree-2  $M_0$-vertex with two $U$-neighbors $v$ and $v'$.
Then  any $(M,I)$-dim $F$ in $G$ satisfies $|V(F)\cap \{v, v' \}|=1$.}
\pff{Since $u$ is in $M_0$ and has no $M_0$-neighbor, exactly one of the $U$-neighbors of $u$
needs to be an end-point of any $(M,I)$-dim $F$; i.e.,  $|V(F)\cap \{v, v'\}|=|(V(G)\setminus V(F))\cap \{v, v'\}|=1$.
}\medskip

From \refl{remove_edge} and \refl{tri1}, we get the following reduction rule.

\rul{rule_edge1}{Let $u$ be a degree-2  $M_0$-vertex with  two $U$-neighbors $v$ and $v'$.
If there is an edge $vv'$, remove the edge $vv'$ from the graph.}

We can use \refl{remove_edge} and \refl{tri1} to prove the correctness of this rule.
Let $(G',M,I)$ be the instance after deleting the edge $vv'$ from $(G,M,I)$ by applying \refr{rule_edge1}.
By \refl{remove_edge} and \refl{tri1}, we have that $(G,M,I)$ is an yes-instance
 if and only if so is $(G',M,I)$.
%

\llem{fivecycle}{
Let   $(G,M,I)$ be a pseudo-feasible  instance with $I=M_1=\emptyset$,
and   let  $u$ be a degree-2   $M_0$-vertex with  two $U$-neighbors $v_1$ and $v'_1$.
If $G$ has  two 2-paths  $u v_1v_2$ and $u v'_1 v'_2$ with $v_2\neq v'_2$,
then  any $(M,I)$-dim $F$ in $G$ satisfies $|V(F)\cap \{v_2,v'_2\}|=1$. }

\pff{By \refl{i-reducible}(a), each of $v_1$ and $v'_1$ has no $M_0$-neighbor other than $u$,
and it holds $\{v_1,v_2,v'_1,v'_2\}\subseteq U$.
Let $F$ be an arbitrary $(M,I)$-dim.
By \refl{tri1}, it holds that  $|V(F)\cap \{v_1,v'_1\}|=1$.
Also we see that $|V(F)\cap \{v_1,v_2\}|=|V(F)\cap \{v'_1,v'_2\}|=1$ for edges $v_1v_2$ and $v'_1v'_2$.
Hence $|V(F)\cap \{v_2,v'_2\}|=1$. }\medskip

From \refl{remove_edge} and \refl{fivecycle}, we get the following reduction rule
to deal with some $M_0$-vertices contained in 5-cycles.

\rul{rule_5cycle}{Let $u$ be  a degree-2  $M_0$-vertex with  two $U$-neighbors  $v_1$ and $v'_1$.
If $G$  has a 5-cycle  $u v_1 v_2 v'_2 v'_1$,  then remove the edge $v_2 v'_2$ from the graph.}

\llem{sixcycle}{Let $v_1v_2v_3v_4v_5v_6$
be a 6-cycle in a pseudo-feasible instance $(G,M,I)$   with $I=M_1=\emptyset$ such that $v_1,v_2$ and $v_3$ are degree-2 vertices
and $v_2$ is
an $M_0$-vertex, where $v_5$ is also an $M_0$-vertex since the instance is pseudo-feasible.
Then the  instance $\mathcal{I}=(G,M,I)$ is an yes-instance
if and only if the instance $\mathcal{I}'=(G\setminus\{v_1,v_2,v_3\},M\setminus\{v_2\},I\cup(N(v_5)\setminus\{v_4,v_6\}))$ is an yes-instance.}
\pff{
First, we show that if the instance $\mathcal{I}$ admits an $(M,I)$-dim $F$ then $F'=F\setminus\{v_1v_2,v_2v_3\}$ is an $(M\setminus\{v_2\},I\cup(N(v_5)\setminus\{v_4,v_6\}))$-dim in
$\mathcal{I}'$. Since the $M_0$-vertex $v_2$ is adjacent to only two $U$-vertices $v_1$ and $v_3$, we know that exactly one of $v_1v_2$ and $v_2v_3$ is in $F$.
When $v_1v_2$ (resp., $v_2v_3$) is in $F$, then
neither of $v_3$ and $v_6$ (resp., $v_1$ and $v_4$) is in $V(F)$ and $v_4v_5$ (resp., $v_5v_6$) is in $F$.
Then we see that $F\setminus\{v_1v_2\}$ (resp.,  $F\setminus\{v_2v_3\}$) is an $(M\setminus\{v_2\},I\cup(N(v_5)\setminus\{v_4,v_6\}))$-dim in
$\mathcal{I}'$.

Second, we show that if $\mathcal{I}'$ admits an $(M\setminus\{v_2\},I\cup(N(v_5)\setminus\{v_4,v_6\}))$-dim $F'$ then either $F'\cup\{v_1v_2\}$ or $F'\cup\{v_2v_3\}$ is
an $(M,I)$-dim in $\mathcal{I}$. In $\mathcal{I}'$, $v_5$ is an $M_0$-vertex adjacent to only two $U$-vertices $v_4$ and $v_6$.
We know that exactly one of $v_4v_5$ and $v_5v_6$ is in $F'$. When $v_4v_5$ (resp., $v_5v_6$) is in $F$,
then neither of $v_3$ and $v_6$ (resp., $v_1$ and $v_4$) is in $V(F')$.
Hence $F'\cup \{v_1v_2\}$ (resp.,  $F'\cup \{v_2v_3\}$) is an $(M,I)$-dim in
$\mathcal{I}$.
 }\medskip

From the above lemma and its proof, we get the following reduction rule.
\reff{reductions}(a) illustrates the operation of the reduction rule.

\rul{rule_6cycle}{
Let $v_1v_2v_3v_4v_5v_6$ be a 6-cycle in a pseudo-feasible instance $(G,M,I)$   with $I=M_1=\emptyset$.
If $v_1,v_2$ and $v_3$ are degree-2 vertices and $v_2$ is
an $M_0$-vertex, then delete $v_1,v_2$ and $v_3$ from the graph and move  to $I$ the vertices in $N(v_5)\setminus\{v_4,v_6\}$.
 For the weighted versions, also update the weight of $v_4v_5$ and $v_5v_6$ by letting $w(v_4v_5)\leftarrow w(v_4v_5)+w(v_1v_2)$ and
$w(v_5v_6)\leftarrow w(v_5v_6)+w(v_2v_3)$.}

\vspace{-0mm}\fig{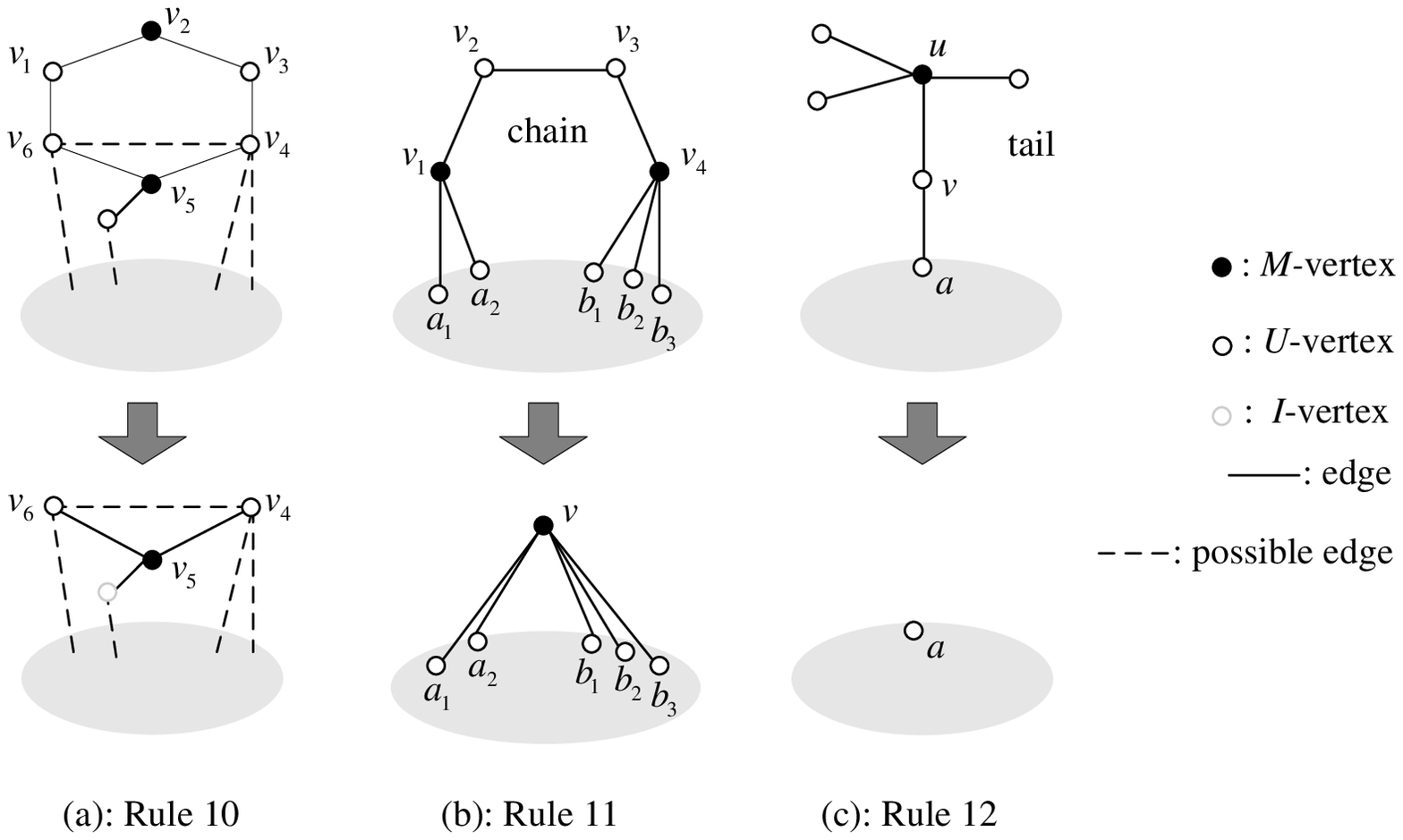}{1}{Reduction operations for Rules~10, 11 and 12}{reductions}\vspace{-0mm}

In a pseudo-feasible  instance $(G,M,I)$  with $I=M_1=\emptyset$,
a 3-path $v_1v_2v_3v_4$ is called a \emph{chain}
if $v_2$ and $v_3$ are degree-2 vertices and $v_1$ is an $M_0$-vertex.
Note that $v_4$ is also an $M_0$-vertex and   no vertex is adjacent to both of $v_1$ and $v_4$
 since the instance is pseudo-feasible.
See \reff{reductions}(b) for an illustration of a chain.

\llem{chain}{Let $v_1v_2v_3v_4$ be a chain in a pseudo-feasible instance $(G,M,I)$   with $I=M_1=\emptyset$.
Let $G'$ be the graph obtained from $G$ by contracting vertices $v_1,v_2,v_3$  and $v_4$
 into a new $M_0$-vertex $v$.
Then  $(G,M,I)$ is an yes-instance
if and only if  $(G', M'=(M\setminus\{v_1,v_4\})\cup\{v\} , I)$ is an yes-instance.
}
\pff{
First, we show that if $(G,M,I)$ admits an $(M,I)$-dim $F$ then $F'=F\setminus\{v_1v_2,v_3v_4\}$ is an $(M',I)$-dim in
$(G',M',I)$.
%
Since  $\{v_1,v_4\}\subseteq M_0$ and $v_2$ and $v_3$ are degree-2 vertices,
we know that exactly one of $v_1v_2$ and $v_3v_4$ is in $F$.
When $v_1v_2$ (resp., $v_3v_4$) is in $F$,
 $v_3$ and $N(v_1)\setminus\{v_2\}$ (resp., $v_2$ and $N(v_4)\setminus\{v_3\}$) are not in $V(F)$.
Then we see that $F\setminus\{v_1v_2\}$ (resp.,  $F\setminus\{v_3v_4\}$)
is an $((M\setminus\{v_1,v_4\})\cup\{v\},I)$-dim in $G'$.

Second, we show that if $(G',M\cup\{v\}\setminus\{v_1,v_4\},I)$ admits an $(M\cup\{v\}\setminus\{v_1,v_4\},I)$-dim $F'$ then either $F'\cup\{v_1v_2\}$ or $F'\cup\{v_3v_4\}$ is
an $(M,I)$-dim in $(G,M,I)$.
In $(G',M',I)$,
an edge $vc$ incident on $v$ is in $F'$ since $v$ is an $M_0$-vertex.
Note that $c$ is adjacent to exactly one of  $v_1$ and $v_4$.
If $c$ is a neighbor of $v_4$ in $G$, then $F'\cup\{v_1v_2\}$ is an $(M,I)$-dim in  $G$.
On the other hand, $c$ is a neighbor of $v_1$ in $G$, and $F'\cup\{v_3v_4\}$ is
an $(M,I)$-dim in   $G$.
}

%

\rul{rule_chain}{
For a chain  $v_1v_2v_3v_4$, contract the four vertices $v_1,v_2,v_3$  and $v_4$
 into a new $M_0$-vertex $v$,
as shown in \reff{reductions}(b).
For the weighted versions, also update the weight of each edge incident on $v$
by letting $w(va)\leftarrow w(va)+w(v_3v_4)$, $a\in N(v_1)\setminus \{v_2\}$ and
 $w(vb)\leftarrow w(vb)+w(v_1v_2)$, $b\in N(v_4)\setminus \{v_3\}$.}

In a pseudo-feasible  instance $(G,M,I)$  with $I=M_1=\emptyset$,
the induced subgraph $G[N[u]]$ for an $M_0$-vertex $u$ is called  a \emph{tail}
if there is exactly one edge between $N[u]$ and $N_2(u)$.
See \reff{reductions}(c) for an illustration of a tail.

\llem{reduction1}{
Let $N[u]$ be a tail in a pseudo-feasible instance $(G,M,I)$ with $I=M_1=\emptyset$.
Let $G'$ be the graph obtained from $G$ by deleting $N[u]$.
Then  $(G,M,I)$ is an yes-instance
if and only if so is $(G', M'=M\setminus \{u\}, I)$.}
\pff{
Assume that the unique edge between $N[u]$ and $N_2(u)$ is $va$, where $v\in N(u)$ and $a\in N_2(u)$.

If $(G,M,I)$ admits an $(M,I)$-dim $F$, then $F$ contains exactly one edge $e$ incident on $u$. It is easy to see that
$F=F'-\{e\}$ is an $(M',I)$-dim in $G'$. If $(G',M',I)$ admits an $(M',I)$-dim $F'$ then $F=F'\cup\{uv^*\}$ is an $(M,I)$-dim in $G$, where $v^*=v$ if $a\not \in V(F')$ and $v^*\in N(u)\setminus \{v\}$ if
$a\in V(F')$.

Therefore $(G,M,I)$ admits an $(M,I)$-dim if and only if $(G',M',I)$ admits an $(M',I)$-dim. }

\rul{rule_tails}{Remove any tail $N[u]$ in a pseudo-feasible  instance $(G,M,I)$,
as shown in \reff{reductions}(c).
For the weighted versions, also update the weight of each edge $e$ incident on $u$
by letting $w(e)\leftarrow w(e)+w(uv_0)-w(uv)$, where  $uv_0\neq uv$ is an edge of minimum weight incident on $u$ except $uv$.
}

\rul{rule_small}{Let $(G,M,I)$ be a pseudo-feasible  instance with $I=M_1=\emptyset$.
If $V(G)=N[u]$ for an $M_0$-vertex $u$, then
move a $U$-neighbor $v$ of $u$ to $M$, where
$v$ is an arbitrary $U$-neighbor of $u$ for the unweighted version whereas $v$ is chosen
so that the edge $vu$ has the minimum weight among all edges incident on $u$
for the minimum version of the problem.
}

An instance is \emph{reduced} if none of   \refr{rule0}-\refr{rule_small} can be applied.

\llem{reduced}{Let $(G,M,I)$   be a reduced instance with $I=M_1=\emptyset$ such that $G$ is connected.
Then \\
{\rm (i)}
Every degree-1 vertex is in $U$ and its unique neighbor is in $M_0$; \\
{\rm (ii)} No $M_0$-vertex is in a triangle or a 4-cycle or has a common $U$-neighbor with other $M_0$-vertex.
No degree-2 $M_0$-vertex is  in a 5-cycle; and \\
{\rm (iii)} Every $M_0$-vertex $u$ has at least two $U$-neighbors and at least two edges between $N(u)$ and
$N_2(u)$.
}
\pff{
{\rm (i)} If a degree-1 vertex $v$ is in $V(G)\setminus U=M_0$, then $N[v]$ would be in $M_1$ by \refr{rule3}.
If the unique neighbor of a degree-1 $U$-vertex is in $U$, then it would be m-reducible by \refl{m-reducible}(a).

{\rm (ii)} Clearly any common $U$-neighbor of some two $M_0$-vertices is i-reducible.
 If $u\in M_0$ is in a triangle, then
 \refr{rule4} (by \refl{i-reducible2}) or \refr{rule_edge1} can be applied.
If $u\in M_0$ is in a 4-cycle, then \refr{rule4} (by \refl{i-reducible2}) can be applied.
By \refr{rule_5cycle}, no degree-2 $M_0$-vertex is contained in a 5-cycle.

{\rm (iii)}
If the number of $U$-neighbors of an $M_0$-vertex $u$ is 0 (resp., 1), then
 \refr{rule0} (resp., \refr{rule4})   can be applied.
If there is at most one edge between $N(u)$ and $N_2(u)$,
 then
 \refr{rule_tails} or \refr{rule_small} can be applied.
}

\section{Ideas for Design and Analysis of Algorithm}\label{sec_alg}
Our algorithm is simple in the sense that it always branches on a $U$-vertex $v$ in a reduced instance by moving it to either $M$ or $I$.
In what follows, we make a basic analysis on how many $U$-vertices will be moved to $M\cup I$ in each of the two branches,
and then define a ``good'' vertex to branch on so that the number of $U$-vertices efficiently decreases in each of the resulting instances.
Recall that $I=M_1=\emptyset$ and $G$ is connected.


In our algorithm, there is only one step where $M=\emptyset$ may hold.
In this step, we branch on  a $U$-vertex by moving it to either $M$ or $I$, which is easy to analyze.
For the other steps, $M_0$ is not empty and the algorithm will branch on a  $U$-vertex $v$ adjacent to an $M_0$-vertex $u$.
In the following, we mainly assume that $v$ is
adjacent to an $M_0$-vertex $u$, where $N(u)$ and $N(v)$ have no common vertex and no edge between them
since no $M_0$-vertex is in a triangle or 4-cycle by \refl{reduced}(ii).
We distinguish (I) the first branch of moving  $v$  to $M$ and (II) the second branch of moving  $v$  to $I$
in a reduced instance $(G,M=M_0\cup M_2,I=\emptyset)$.

(I) The first branch of moving  $v$  to $M$:
Then all vertices in $N(uv)$, which are all $U$-vertices,
 will be moved to $I$, and all vertices in $N_2(uv)\cap U$ will be moved to $M$ directly by applying our reduction rules.
So in this branch, the number of $U$-vertices decreases by at least
$$1+|N(uv)|+|N_2(uv)\cap U|.$$
We also analyze two special cases where  some  $U$-vertices in $V\setminus N_2[uv]$ can be eliminated. \\
\noindent
- The first special case is that there are two vertices $z,z'\in N_2(uv)$ adjacent to each other,
 including the case that $u$ is contained in a 5-cycle or 6-cycle.
Any vertex $w\in N_3(uv)$ adjacent to one of $z$ or $z'$ is
 a $U$-vertex, since otherwise $v$ would be i-reducible.
Such a vertex $w$ will be eliminated by \refr{rule1}, since $z$ and $z'$ will be a pair of adjacent $M_1$-vertices after moving $v$ to $M$.
 \\
\noindent - The second special case is that there is a vertex $z\in N_2(uv)$ adjacent to exactly
one vertex $w\in N_3(uv)$, where we can assume that $z$ has no  other  $N_2(uv)$-neighbor
 since otherwise  the first case can be applied.
The vertex $w$ is also a $U$-vertex since otherwise $v$ would be i-reducible.
We can see that $w$ will be moved to $M$ directly by \refr{rule3}.

Hence for each of the two special cases we can decrease one more $U$-vertex $w$.
Moreover any other $U$-neighbor $w'$ of $w$ will also be eliminated by the reduction rules.
In the following sections,  we  sometimes prove the existence of such a vertex $w'$
to ensure that two $U$-vertices $w$ and $w'$ can be further eliminated in the first branch.

(II) The second branch of moving  $v$  to $I$:
Note that $N(v)\setminus\{u\}\subseteq U$ since $M_0$-vertex $u$ has no common $U$-vertex with any other $M_0$-vertex by \refl{reduced}(ii).
In this branch all vertices in $N(v)\setminus\{u\}~(\subseteq U)$ will be moved to $M$, and the number of $U$-vertices decreases by at least
$$1+|N(v)\setminus\{u\}|.$$
In a reduced instance with a non-empty set $M_0$, we can always find
a $U$-vertex $v$ with  an $M_0$-neighbor $u\in N(v)$ and a $U$-neighbor $v'\in N(v)$ by \refl{reduced}(iii).
Such a $U$-vertex $v$ satisfies
$|N(v)\setminus\{u\}|\geq 1$.
So in the second branch, the number of $U$-vertices decreases by at least 2.
We show two special cases where  more $U$-vertices in $V\setminus N[v]$ can be eliminated.\\
\noindent -
The first special case is that $v$ has a degree-2 $U$-neighbor $v'$, where  we denote $N(v')=\{v,v''\}$.
After $v$ is moved to $I$, both of $v'$ and $v''$ will be moved to $M$, and any other vertices adjacent to $v''$ will be moved to $I$. \\
\noindent - The second special case is that $u$ is of degree 2, where  we denote $N(u)=\{v,v'\}$.
 After $v$ is moved to $I$, $v'$ will be moved to $M$.
Thus the second branch is equivalent to the operation of moving $v'$ to $M$,
and we see that  the $U$-vertices in $\{v'\}\cup N(v'u) \cup (N_2(v'u)\cap U)$ will be eliminated.

We are ready to define ``good'' vertices to branch on.
 A $U$-vertex  is called an \emph{effective} vertex if it is of degree at least 3 and has an $M_0$-neighbor.

Our algorithm branches on vertices as follows.
We first select effective vertices to branch on as long as they exist.
After this step, no effective vertex exists and each $U$-neighbor of an $M_0$-vertex is of degree at most 2.
Next our algorithm tests whether the graph contains an $M_0$-vertex $u$
having only two degree-2 $U$-neighbors,
and selects a degree-2 $U$-neighbor $v$ of such an $M_0$-vertex $u$ to branch on, if any.
Note that every degree-2 $M_0$-vertex has two degree-2 $U$-neighbors, since otherwise it would be in a tail.
After this, no $M_0$-vertices of degree 2 exist  any more.
Then our algorithm tests whether the graph has an $M_0$-vertex contained in 5-cycles
 and selects one neighbor $v$ of such an $M_0$-vertex to branch on, if any.
Finally no $M_0$-vertex is contained in a cycle of length 3, 4 or 5,
and our algorithm selects a neighbor $v$ maximum $|N_2(uv_1)|$ of any $M_0$-vertex $u$ or
a $U$-vertex $v$ with some other priority (if $M_0=\emptyset$) to branch on.
The main steps of our algorithm are described in Figure~\ref{dim}.
Note that Step~3 is based on \refl{pseudo}.
Then it is easy to observe the correctness of the algorithm.

 \vspace{-0mm}\begin{figure}[!h] \setbox4=\vbox{\hsize28pc
\noindent\strut
\begin{quote}
\vspace*{-5mm}
\textbf{Input}: An instance $(G,M,I)$ of a graph $G=(V,E)$ and two subsets of vertices $M$ and $I\subseteq V$,
where initially $M=I =\emptyset$.
\\
\textbf{Output}: 1 if an $(M,I)$-dim exists in $G$ and 0 otherwise.\\
\vspace*{-2mm}
\begin{enumerate}
\item \textbf{If}
   \{$|U|\leq 6$ or the maximum degree of the graph is at most 2\},
     Solve the instance directly and return 1 if an $(M,I)$-dim exists and 0 otherwise.
\item \textbf{Elseif}
    \{The instance is not pseudo-feasible\},
      Apply  \refr{rule0}-\refr{rule5} until none of them can be applied any more.
\item \textbf{Elseif}
   \{$I\cup M_1\neq\emptyset$ or $G[M_0\cup U]$ contains more than one component\},
     Let $H_1,H_2,\ldots,H_p$ be the components of $G[M_0\cup U]$;
     Return  $\mathrm{dim}(H_1, M_0\cap V(H_1), \emptyset)\wedge\mathrm{dim}(H_2, M_0\cap V(H_2), \emptyset)
    \wedge \cdots \wedge \mathrm{dim}(H_p, M_0\cap V(H_p), \emptyset)$.\\
     /* After Step 3, it always holds $I\cup M_1=\emptyset$. */
\item \textbf{Elseif}
    \{The instance is not a reduced instance\}, Apply  one of \refr{rule_edge1}-\refr{rule_small} in the listed order
 and return $\mathrm{dim}(G',M',I')$, where $(G',M',I')$ is the resulting instance after applying the rule.
\item \textbf{Else}
    Choose a $U$-vertex $v_1$ as follows and branch on it
by returning $\mathrm{dim}(G, M\cup\{v_1\},I)\vee\mathrm{dim}(G, M,I\cup\{v_1\})$:
\begin{enumerate}
\item \textbf{If}\{There is some effective vertex\},
Choose an effective vertex $v_1$ adjacent to an $M_0$-vertex $u$ so that the degree of $u$ is minimized.
\item \textbf{Elseif}
\{There is a degree-2 $M_0$-vertex\}, Choose a degree-2 $M_0$-vertex $u$ and let $v_1$ be a neighbor of $u$.
\item
\textbf{Elseif}\{There is an $M_0$-vertex with exactly two degree-2 $U$-neighbors\},
Choose such an $M_0$-vertex $u$, and
 let $v_1$ be  a degree-2 neighbor of $u$ with maximum $|N_2(uv_1)|$.
\item
\textbf{Elseif}\{There is an $M_0$-vertex $u$ that is in a 5-cycle $uv_2a_1a_2v_3$
 and  has at least three degree-2  neighbors\},
Choose such a 5-cycle $uv_2a_1a_2v_3$, and let $v_1$ be a degree-2 vertex in $N(u)\setminus\{v_2,v_3\}$.
\item
\textbf{Elseif} \{There is still an $M_0$-vertex $u$\}, Let $v_1$ be a neighbor of $u$ with maximum $|N_2(uv_1)|$.
\item
\textbf{Else} /*$M_0=\emptyset$ */
Let $v_1$ be a $U$-vertex satisfying one of the following: (i) $v_1$ is contained in a triangle or
 4-cycle;
(ii) no such vertices in (i) exist, and $v_1$ is adjacent to at least one degree-2
 vertex; and (iii) no such vertices in (i) and (ii) exist, and $v_1$ is of maximum degree.

\end{enumerate}
\end{enumerate}

\end{quote} \vspace*{-5mm} \strut}  $$\boxit{\box4}$$ \vspace*{-2mm}
\caption{Algorithm $\mathrm{dim}(G,M,I)$} \label{dim} \vspace{-0mm}
\end{figure}

\section{The Detailed Analysis}\label{sec_analy}

Only Step~5 in the algorithm creates recurrences. Before analyzing each substeps in Step~5,
we prove some properties of reduction operations.

%

\llem{reductionu}{After applying any step of $\mathrm{dim}(G,M,I)$, the total number of $U$-vertices in the instance does not increase.}

No operation in the algorithm will create any new $U$-vertex. It is easy to see the correctness of this lemma.

\llem{Mconnected}{Let $(G,M,I)$ be an instance such that $G$ is a connected graph and $M\neq \emptyset$.
Assume that  applying a  branching rule or reduction rule except \refr{rule_tails} to $(G,M,I)$
 results in an instance $(G',M',I')$ without solving the instance directly.
Then   $G'$ is still a connected graph and $M'\neq \emptyset$.
}
\pff{In \refr{rule0} to \refr{rule5}, \refr{rule_small} and the branching operations in Step~5 of the algorithm, either the problem is solved directly or some $U$-vertices are moved to $M\cup I$.
In the latter case, the connectivity of the graph is not affected and no $M$-vertices are removed from the graph.
In \refr{rule_edge1} and \refr{rule_5cycle}, one edge is removed from the graph, where the graph remains connected
 since we see that this edge is not a bridge in the graph.
In \refr{rule_6cycle}, some vertices including an $M_0$-vertex are removed from the graph,
where the graph remains connected and still has an $M$-vertex.
\refr{rule_chain} contracts
some vertices without disconnecting the graph and keeping at least one $M$-vertex in the remaining graph.
This proves the lemma.
}

By \refl{Mconnected} and \refc{cut}, we know that if a pseudo-feasible instance has the property that each connected component of $G[U\cup M_0]$ contains at least one $M_0$-vertex,
then the resulting pseudo-feasible instance still satisfies this property
after applying any branching rule or reduction rule except \refr{rule_tails}.
 This will be used in the analysis in Step~5(f).

Next we give the detailed analysis of each substep  in Step~5.
When a vertex $v_1$ is chosen in Step~5,
let  $\Delta_M$ (resp., $\Delta_I$) denote the number   of
$U$-vertices that decrease by the branch of moving $v_1$ to $M$ (resp., to $I$)
and by possible applications of reduction rules to the resulting instance.

\subsection{Step~5(a)}
For an instance in Step~5(a),  the graph has some effective vertex.
Let $v_1$ be an effective vertex adjacent to an $M_0$-vertex $u$ and at least two $U$-vertices $a_1,a_2\in N_2(u)$.
We assume that the algorithm will branch on $v_1$.

First we show that $|N(uv_1)|\geq 3$ and $|N_2(uv_1)|\geq 2$.
Since $u$ is of degree at least 2 by \refl{reduced} and has another $U$-neighbor $v_2 \neq v_1$, we have $|N(uv_1)|\geq 3$.
Now cnsider the vertices in $N_2(uv_1)$.
Note that no pair of vertices in $N(uv_1)$ are adjacent to each other,
since otherwise $v_1$ would be i-reducible.
Neither of $a_1$ and $a_2$ can be a degree-1 vertex, since otherwise $v_1$ would be in $M$.
Then each of $a_1$ and $a_2$ is adjacent to a vertex in $N_2(uv_1)$.
If there is only one vertex $c$ in  $N_2(uv_1)$, then
$a_1$ and $a_2$ are adjacent to the same vertex $c$, and $v_1$ would be m-reducible.
Therefore we have that each of $a_1$ and $a_2$  has  an $N_2(uv_1)$-neighbor,  and it holds $|N_2(uv_1)|\geq 2$.

Next we derive an upper bound on  $\Delta_M$ (resp., $\Delta_I$).
We define $\lambda(u)=\min\{1, |N(u)|-2\}$; i.e.,  $\lambda(u)= 0$ if $u$ is of degree 2
 and $\lambda(u)=1$ if $u$ is of degree $\geq 3$.
Let $x=|N_2(uv_1)\cap U|$.

In the branch of moving $v_1$ to $M$,  all vertices in $N(uv_1)$
will be moved to $I$ and all vertices in $N_2(uv_1)$ will be moved to $M$.
In this branch, the number of $U$-vertices decreases by at least $1+|N(uv_1)|+x$.
Note that $N(uv_1)$ contains at least three vertices $v_2, a_1$ and $a_2$,
and when $u$ is  of degree $\geq3$ it holds $|N(uv_1)|\geq 4$.
Then  $\Delta_M\geq 1+|N(uv_1)|+x\geq 4+\lambda(u)+x$.

For the other branch of moving $v_1$ to $I$, we prove that
$\Delta_I\geq 8-\lambda(u)-\min\{x,4\} $
 by distinguishing five different values of $x$.

Case 1. $x=0$: In this case, $N_2(uv_1)$ contains at least two vertices $c_1,c_2\in M_0$ since $|N_2(uv_1)|\geq 2$.
We assume that $c_1$ (resp., $c_2$) is adjacent to $a_1$ (resp., $a_2$),
where  $c_1$ and $c_2$ are not adjacent to each other, since otherwise they would be in $M_0$.
Since  no two vertices in $M$  have a common $U$-neighbor in a reduced instance by \refl{reduced},
we see that $c_1$ and $c_2$ have different $N_3(uv_1)$-neighbors:  $c_1$ is adjacent to say, $c_1'\in N_3(uv_1)$
 and $c_2$ is adjacent to say, $c_2'\in N_3(uv_1)$ such that $c_1'\neq c_2'$.
 \reff{case1} illustrates the neighbors of edge $uv_1$.
We easily see that there is no edge between $\{a_1,c_1,c_1'\}$ and $\{a_2,c_2,c_2'\}$, since otherwise
 the graph would have  some i-reducible or m-reducible vertex.
Since $N[c_1]$ and $N[c_2]$ are not tails, we know that there are two vertices $w,w'\in U\setminus \{u,v_1,v_2,a_1,a_2,c_1,c_2,c_1',c_2'\}$ such that either
(i) each of $w$ and $w'$ is adjacent to a vertex in $\{c_1,c_2,c_1',c_2'\}$ or
(ii) $w$ is adjacent to both of $c'_1$ and $c'_2$ and a vertex $w''$ is adjacent to $w$.
For the latter case, $w''$ is possibly an $M$-vertex.
Since $w$ is not an $M$-vertex, $w''$ cannot be
a degree-1 vertex, and thereby it has a $U$-neighbor $w'$.
In any of (i) and (ii), when $v_1$ is moved to $I$, at least seven $U$-vertices $\{v_1,a_1,a_2,c_1',c_2',w,w'\}$ will be eliminated from $U$, implying that $\Delta_I\geq 7=8-\lambda(u)-\min\{x,4\}$ when $\lambda(u)=1$.
Furthermore, when $u$ is of degree 2, the other neighbor $v_2$ of $u$ will also be moved to $I$ and
then at least eight $U$-vertices $\{v_1,v_2,a_1,a_2,c_1',c_2',w,w'\}$ will be eliminated from $U$,
 implying that $\Delta_I\geq 8=8-\lambda(u) -\min\{x,4\}$.

\vspace{-0mm}\fig{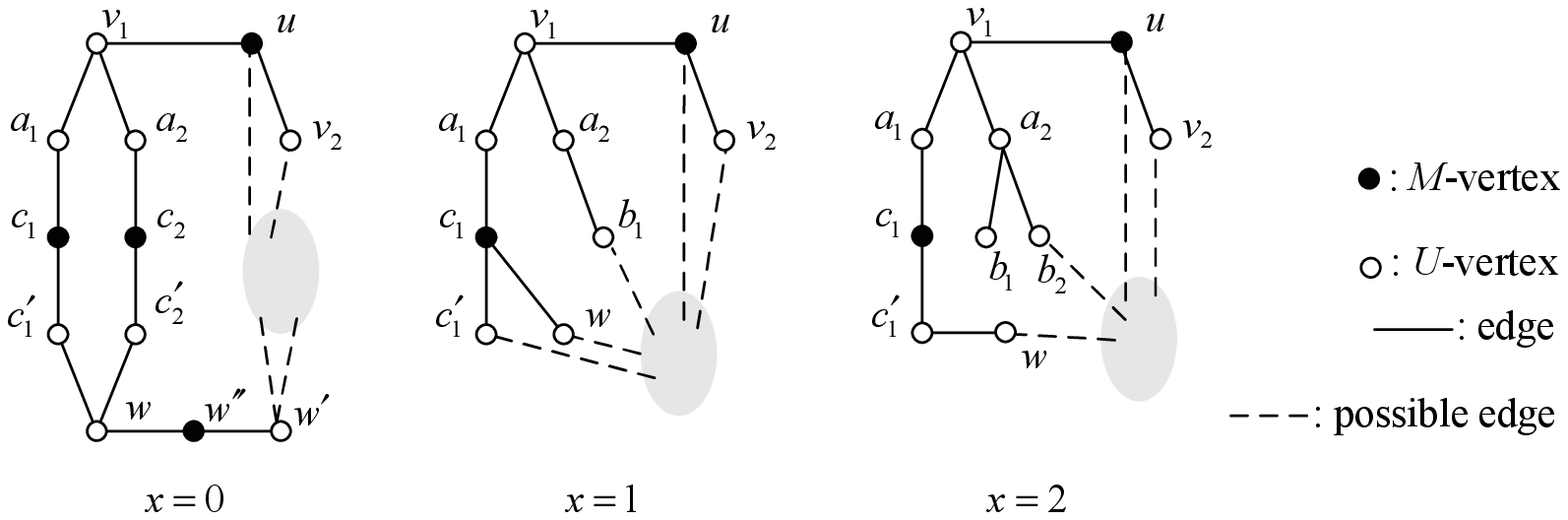}{1}{Neighbors of edge $uv_1$ in an instance at Step~5(a)}{case1}\vspace{-0mm}

Case 2. $x=1$: In this case,  $N_2(uv_1)$ contains at least one vertex $c_1\in M_0$ and one vertex $b_1\in U$.
Note that $c_1$ is not in $N_2(u)$ because no two $M$-vertices have a common $U$-neighbor.
Then $c_1$ is adjacent to some neighbor of $v_1$. Without loss of generality, we can assume that
$c_1$ (resp.,   $b_1$)  is adjacent to $a_1$  (resp.,  $a_2$).
Furthermore, $c_1$ and $b_1$ cannot be adjacent to each other, since otherwise $v_1$ would be u-reducible.
See \reff{case1} for an illustration of  the neighbors of edge $uv_1$.
 Analogously with the above case, we know that $c_1$ is adjacent to a vertex $c_1'\in N_3(uv_1)$.
Let $V'=\{u,v_1,v_2,a_1,a_2,c_1,b_1,c_1'\}$.
There is also a vertex $w\in U\setminus V'$ adjacent to either $c_1$ or $c_1'$.
When $v_1$ is moved to $I$, at least six vertices $\{v_1,a_1,a_2,b_1,c_1',w\}$ will be eliminated from $U$,
implying that $\Delta_I\geq 6=8-\lambda(u)-\min\{x,4\}$ when $\lambda(u)=1$.
Furthermore, when $u$ is of degree 2, the other neighbor $v_2$ of $u$ will also be moved to $I$ and
then at least seven $U$-vertices $\{v_1,v_2,a_1,a_2,b_1,c_1',w\}$ will be eliminated from $U$,
 implying that $\Delta_I\geq 7=8-\lambda(u)-\min\{x,4\}$.

Case 3. $x=2$: Let $\{b_1,b_2\}=N_2(uv_1)\cap U$.
In the branch where $v_1$ is moved to $I$, three $U$-vertices $\{v_1, a_1, a_2\}$ will be eliminated. If $u$ is of degree 2, the other neighbor $v_2$ of $u$ will also be eliminated. We show that at least two more $U$-vertices will be eliminated in this branch,
which implies that $\Delta_I\geq 3+2=8-\lambda(u)-\min\{x,4\}$ for $\lambda(u)=1$ and
$\Delta_I\geq 3+1+2=8-\lambda(u)-\min\{x,4\}$ for $\lambda(u)=0$.
We distinguish  two subcases.

(i) $b_1$ and $b_2$ have a common neighbor in $N(v_1)$:
See \reff{case1}.
If no vertex in $N(v_1)\setminus \{u\}$  has a neighbor in $N_2(v_1)\cap M_0$, then $v_1$ would be m-reducible.
Then at least one vertex in $N(v_1)\setminus \{u\}$, say $a_1$ has a neighbor $c_1\in N_2(v_1)\cap M_0$.
Since $a_1$ is not in $M$ and $c_1$ cannot be a degree-1 vertex, we know that $c_1$ has a $U$-neighbor $c_1'\in N_3(v_1)$.
Note that path $c_1'c_1a_1$ cannot be a tail. Then $c_1$ or $a_1$ should have a $U$-neighbor $w\in \{v_1,v_2,a_1,a_2,c_1'\}$ and it will be moved to $I$ or $M$ after $v_1$ is moved to $I$.
Then at least two more $U$-vertices, $c_1'$ and $w$, will be eliminated from $U$.

(ii) $b_1$ and $b_2$  have no common neighbor in $N(v_1)$:
Then each of $a_1$ and $a_2$ is adjacent to at most one vertex in $N_2(uv_1)\cap U$. If one of $a_1$ and $a_2$, say $a_1$ is not adjacent to any vertices in $N_2(uv_1)\cap U$, we can assume
that $a_1$ is adjacent to a vertex $c_1\in N_2(v_1)\cap M_0$. From the analysis in Case 3(i), we  see that at least two more $U$-vertices $c_1'$ and $w$ will be eliminated. Next we assume
that each of $a_1$ and $a_2$ is adjacent to exactly one vertex in $N_2(uv_1)\cap U$. Without loss of generality, we assume that $b_1$ (resp., $b_2$) is adjacent to $a_1$  (resp.,  $a_2$).
We   see that: for each $i\in\{1,2\}$, $b_i$ will be moved to $I$ if $a_i$ has a  $(N_2(v_1)\setminus\{b_i\})$-neighbor,
which should be an $M_0$-vertex; and $b_i$ will be moved to $M$ if $a_i$ has no $(N_2(v_1)\setminus\{b_i\})$-neighbor.
Then at least two more $U$-vertices, $b_1$ and $b_2$, will be eliminated from $U$.

Case 4.  $x=3$: At least one of $a_1$ and $a_2$, say $a_1$ has at most  one $N_2(uv_1)\cap U$-neighbor.
If $a_1$ has exactly one $N_2(uv_1)\cap U$-neighbor $b_1$,
then  the four vertices $\{v_1, a_1, a_2,b_1\}$ will be eliminated after $v_1$ is moved to $I$.
Assume that $a_1$ has no $N_2(uv_1)\cap U$-neighbor.
Then  $a_1$ has an $N_2(uv_1)\cap M_0$-neighbor $c_1$, which  must have
 a $(U\setminus\{v_1,v_2, a_1, a_2\})$-neighbor $w$.
For this case, the four vertices $\{v_1, a_1, a_2,w\}$ will be eliminated.
Hence
$\Delta_I\geq 4=8-\lambda(u)-\min\{x,4\}$ for $\lambda(u)=1$.
Note that when $u$ is of degree 2,
the other neighbor $v_2$ of $u$ will also be eliminated in any of the above cases,
 implying that $\Delta_I\geq 4+1=8-\lambda(u)-\min\{x,4\}$ for $\lambda(u)=0$.

Case 5. $x\geq 4$: After moving $v_1$ to $I$, we can always eliminate at least three $U$-vertices $\{v_1, a_1, a_2\}$,
 implying that $\Delta_I\geq 3=8-\lambda(u)-\min\{x,4\}$ for $\lambda(u)=1$.
If $u$ is a degree-2 vertex the other neighbor $v_2$ of $u$ will also be eliminated directly,
 implying that $\Delta_I\geq 3+1=8-\lambda(u)-\min\{x,4\}$ for $\lambda(u)=0$.

\medskip
From the arguments in Cases~1-5, we have that $\Delta_I\geq 8-\lambda(u)-\min\{x,4\}$.
Therefore we can branch with recurrence relations
$$C(n)\leq C(n\!-\!(4+\lambda(u)+x)) \!+\!C(n\!-\!  (8\!-\!\lambda(u)\!-\! \min\{x,4\}) )  \mbox{ for $x\geq 0$}, $$
all of which are covered by
\eqn{2dbranch}{C(n)\leq C(n-(8+\lambda(u)))+C(n-(4-\lambda(u))).}
\medskip

Before we proceed to analysis on Step~5(b),
we analyze a special case in Step~5(a), where the neighbor $u$ of $v_1$
 is an $M_0$-vertex such that all the neighbors are effective vertices.
Although it is covered by the above analysis, we here derive better
recurrence relations for it, which will be used in the analysis of Step~5(e).
The algorithm will branch on $v_1$ by including it to $M$ or $I$ in Step~5(a),
and let $d$ be the degree of $u$ and $N(u)=\{v_1,v_2,\ldots,v_d\}$, which contains only effective $U$-vertices.
We distinguish two cases.

Case S1. $d=2$:
 Then the two $U$-neighbors $v_1$ and $v_2$ of $u$ are  of degree at least 3.
Choose vertices $ a_1,a_2\in N(v_1)\setminus\{u\}$ and $ b_1,b_2\in N(v_2)\setminus\{u\}$.
Then $V'=\{v_1,v_2,a_1,a_2,b_1,b_2\}$ is a set of six different $U$-vertices since $u$ is not contained in any triangle or 4-cycle by \refl{reduced}(ii).
The second branch of moving $v_1$ to $I$ is equivalent to the operation of moving $v_2$ to $M$.
When we move $v_i$ $(i=1,2)$ to $M$, all $U$-vertices in $N_2[v_iu]$ will be eliminated.
We can see that $V'\subseteq N_2[v_1u]\cap N_2[v_2u]$. In each branch, at least six $U$-vertices decrease,
implying that $\Delta_M\geq 6$ and $\Delta_I\geq 6$.
We will further show that $\max\{\Delta_M,\Delta_I\}\geq 7$
to obtain  the following   recurrence relation.
\eqn{6-7}{C(n)\leq C(n-6)+C(n-7).}
For this, we prove that
at least one more $U$-vertex decreases in one of the two instances generated by branching on $v_1$.
Vertices $a_1$ and $a_2$ are not adjacent to each other, since otherwise $v_1$ would be i-reducible.
Each of $a_1$ and $a_2$ must be adjacent to a vertex $N_2(v_1u)\setminus\{b_1,b_2\}$, since otherwise
$v_2$ would be i-reducible.
Assume that $a_1$ (resp., $a_2$)  has a  $(N_2(v_1u)\setminus\{b_1,b_2\})$-neighbor $c_1$ (resp.,  $c_2$).
Let $\alpha$ be the number of $U$-vertices in  $N_2[v_1u]$.
Since   $V'\subseteq N_2[v_1u]$, it holds $\alpha\geq 6$,
where if $\alpha=6$ then both of $v_1$ and $v_2$ are of degree 3
and both of $c_1$ and $c_2$ are $M_0$-vertices.
Note that two $M_0$-vertices $c_1$ and $c_2$ are not adjacent and have no common $U\cup M$-neighbors.
Then each of $c_1$ and $c_2$ has at least two $U$-neighbors, since they cannot be degree-1 vertices.
If one of $c_1$ and $c_2$  has a $U\setminus V'$-neighbor $w$, then in the branch of moving $v_1$ to $I$,
the seven $U$-vertices in $V'\cup \{w\}$ will decrease, implying that $\Delta_I \geq 7$.
Now assume that  neither of $c_1$ and $c_2$ has a $V\setminus V'$-neighbor.
Then
each of $c_1$ and $c_2$ is a degree-2 vertex adjacent to a vertex in $\{b_1,b_2\}$.
At least one $U$-vertex $w\not\in V'$ is adjacent to $b_1$ or $b_2$,
since the graph contains more than six $U$-vertices.
In the branch of moving $v_1$ to $M$, vertices $b_1$ and $b_2$ will become $M_1$-vertices
and $w$ will be included into $I$.
Again
the seven $U$-vertices in $V'\cup \{w\}$ will be eliminated, implying that $\Delta_M \geq 7$.

Case S2. $d\geq 3$:
Note that each vertex in $N(u)$ is of degree at least 3 since it is an effective vertex in the special case.
In the branch where $v_1$ is moved to $M$, all $U$-vertices in $N_2[v_1u]$ are eliminated.
There are at least $2+(d-1)=d+1$ $U$-vertices in $N(v_1u)$. No pair of vertices in $N(v_1u)$ are adjacent otherwise $u$ would be in a triangle or 4-cycle.
Then each vertex in $N(u)\setminus\{v_1\}$ has at least two $N_2(v_1u)$-neighbors,
which are $U$-vertices since otherwise a vertex in $N(u)\setminus\{v_1\}$ would be adjacent to two $M_0$-vertices
and should have been moved to $I$.
Furthermore, no pair of vertices in $N(u)\setminus\{v_1\}$ can share a common neighbor in $N_2(v_1u)$
since otherwise $u$ would be in a 4-cycle.
Then there are at least $2(d-1)$ different $U$-vertices in $N_2(v_1u)$.
In total, the number of $U$-vertices in  $N_2(v_1u)$ is at least $1+(d+1)+2(d-1)=3d$, implying that $\Delta_M \geq 3d$.
In the other branch of moving $v_1$ to $I$, at least three $U$-vertices in $N[v_1]$ decrease, implying that $\Delta_I \geq 3$.
Therefore we get a recurrence relation
\eqn{multi-eff}{C(n)\leq C(n-3d)+C(n-3)  \mbox{  for $d\geq 3$}. }

\medskip
This completes our analysis on recurrence relations  in  Step~5(a).
We also examine  structural property on
instances with no effective vertices.

\llem{withoutopt}{Let $(G,M=M_0,I=\emptyset)$ be a reduced instance having no effective vertices.
For any pair of adjacent vertices $v\in U$ and $u\in M_0$ in $G$, it holds that\\
{\rm (i)}  the degree of $v$ is at most 2; and \\
{\rm (ii)}  $N_2(vu)\subseteq U$.
}
\pff{{\rm (i)} By \refl{reduced}(ii), we know that $u$ is not in a triangle.
Thus, if $v$ is of degree at least 3, then $v$ has  at least two $N_2(u)\cap U$-neighbors and $v$ would  an effective vertices,
 a contradiction to the assumption.

{\rm (ii)} If there is an $M_0$-vertex $u'$ in $N_2(vu)$ then there is a path $uvv'u'$, where $v'\in N(vu)$. If $v'$ is a degree-2 vertex, then $uvv'u'$ is a chain.
 If $v'$ is of degree $\geq 3$, then $v'$ is an effective vertex adjacent to $u'$.
For any case, the instance cannot be a reduced instance having no effective vertex.
}

The  property will be frequently used in the next analysis.

\subsection{Step~5(b) and Step~5(c)}
We  derive recurrence relations to Step~5(b) and Step~5(c).
Recall that
no $M_0$-vertex $u$ is  in any triangle and there are at least two edges between $N(u)$ and $N_2(u)$ by \refl{reduced}.
In Step~5(b),  both of the two neighbors of any $M_0$-vertex $u$ are of degree 2 and
adjacent to some vertices in $N_2(u)$, since no neighbor of $u$ is of degree $\geq3$ by \refl{withoutopt}(i).
Thus, in Step~5(b)-(c), every $M_0$-vertex $u$ has exactly two degree-2 $U$-neighbors,
each of which is adjacent to some vertex in $N_2(u)$,
and any other neighbor of $u$ is of degree 1.

Let $u$ be  a degree-2 $U$-neighbor $v_1$ of an $M_0$-vertex $u$
on which algorithm  branch in Step~5(b) or (c).
Let  $d$ be the degree of $u$ and $N(u)=\{v_1,v_2,\ldots,v_d\}$, where $v_1$ and $v_2$ are of degree 2.
Denote $N(v_1)=\{u,a_1\}$ and $N(v_2)=\{u,a_2\}$, where $a_1\neq a_2$ holds
since otherwise $v_1$ and $v_2$ would be i-reducible.
 Vertex $a_1$ cannot be a degree-2 vertex since otherwise $uv_1a_1$ would be in a chain.
Then $a_1$  has no $N(v_1u)$-neighbor since otherwise $v_1$ would be i-reducible.
 We know that $a_1$ has at least two  $N_2(v_1u)$-neighbors, say  $b_1$ and $b_2$, each of which
 is a $U$-vertex by \refl{withoutopt}(ii).
We distinguish with Step~5(b) and Step~5(c).


Case~1. $d=2$ (Step~5(b)):
Note that $a_2\not\in\{ b_1,b_2\}$, since  $a_2 \in\{ b_1,b_2\}$ would imply
that  edge $a_1a_2$ can be deleted by \refr{rule_5cycle}.
Hence $N_2(v_1u)$ contains at least three different $U$-vertices $a_2, b_1$ and $b_2$.
See \reff{case2} for an illustration of  the neighbors of edge $uv_1$.
In the branch where $v_1$ is moved to $M$,   all vertices in $N(v_1u)$ will be moved to $I$,
 and all vertices in  $N_2(v_1u)$ will be moved to $M$.
The number of  $U$-vertices  eliminated from $\{v_1\}\cup N(v_1u) \cup (N_2(v_1u))$ is at least $1+2+3=6$.
The other branch of moving $v_1$ to $I$ is equivalent to moving $v_2$ to $M$.
By the same argument, we see that this branch  also decreases at least six $U$-vertices.
We get a recurrence relation
\eqn{6-6}{C(n)\leq C(n-6)+C(n-6).}

\vspace{-0mm}\fig{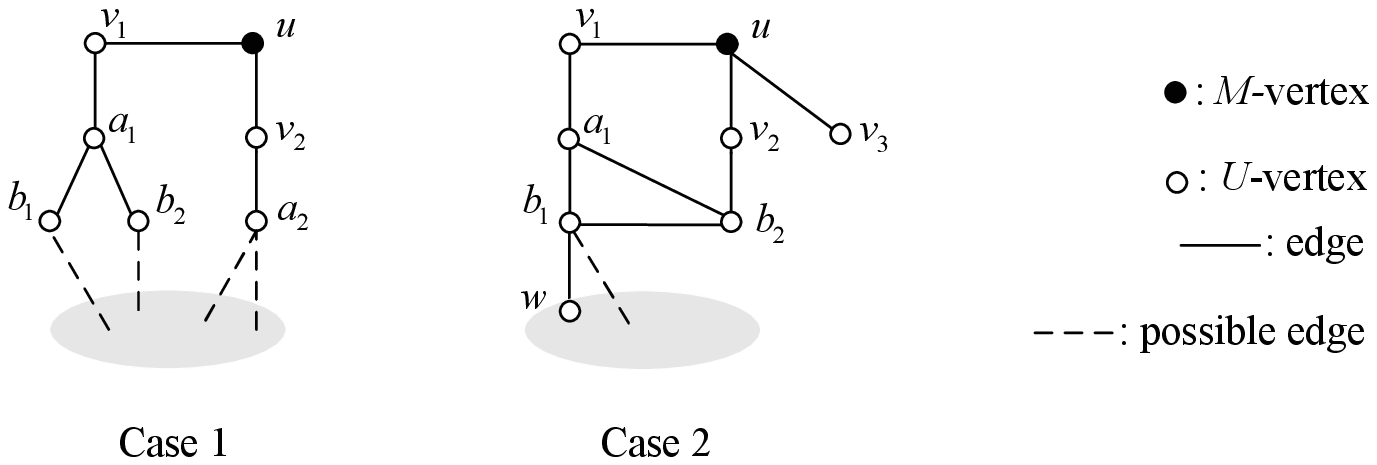}{0.9}{Neighbors of edge $uv_1$ in an instance at Step~5(b) and Step~5(c)}{case2}\vspace{-0mm}

When there is a degree-2 $M_0$-vertex   in the resulting instance,
 the next branch will be in either Step~5(a) or Step~5(b). We have the following lemma.

\llem{degree2}{If a reduced instance $(G,M=M_0, I=\emptyset)$ has a degree-2 $M_0$-vertex,
then the algorithm will branch with a recurrence relation covered by
\eqn{2dbranch_g}{C(n)\leq C(n-8)+C(n-4).}
}
\pff{When there is an effective vertex in the instance, the algorithm branches with a recurrence relation coverted by
 \refe{2dbranch} with $\lambda=0$.
When there is no effective vertex,
the algorithm will execute Step~5(b) for  some degree-2 $M_0$ vertex, and it
  branches with \refe{6-6} by the above analysis, where \refe{6-6} is covered by \refe{2dbranch_g}. }

Case~2. $d\geq 3$ (Step~5(c)): See \reff{case2}.
We show that each of the two branches on $v_1$  decreases at least seven $U$-vertices.
First consider  the branch of moving $v_1$ to $M$ by distinguishing two subcases.

(i) $|N_2(v_1u)|\geq 3$ or $u$ is of degree $\geq 4$:
After $v_1$ in moved to $M$, the number of $U$-vertices decreases by at least
$1+|N(v_1u)|+|N_2(v_1u)|\geq \min\{1+3+3,1+4+2\}=7$.

(ii) $|N_2(v_1u)|=2$ and $u$ is of degree $3$:
Without loss of generality, we assume that $a_2=b_2$, where  $a_2\in N_2(v_1u)$.
For this case, $b_2$ must be a degree-3 vertex,
because if $b_2$ is of degree $\leq 2$ then there would be a chain or tail,
and if $b_2$ is of degree $\geq 4$ then the algorithm must have selected $v_2$ instead of $v_1$ to branch on,
since $|N_2(v_2u)|\geq 3> |N_2(v_1u)|$.
Since  the current graph contains more than six $U$-vertices after Step 1,
  if $b_2$ is adjacent to $b_1$, then $b_1$ is adjacent to  a vertex  $w\in N_3(v_1u)$,
where the vertex $w$ is a $U$-vertex since otherwise $v_1$ would be i-reducible.
On the other hand, if $b_2$ is not adjacent to $b_1$,
 then it must be adjacent to a $U$-vertex $w\in N_3(v_1u)$.
In the branch where $v_1$ in moved to $M$,   all vertices in $N(v_1u)$ will be moved to $I$,
 and all vertices in  $N_2(v_1u)$ will be moved to $M$.
The vertex $w$ in any case will also be moved to $I$ or $M$ directly.
Therefore this branch eliminates at least seven $U$-vertices in $\{v_1,v_2,v_3, a_1, b_1,b_2,w\}$.

Second we consider the other branch where $v_1$ is moved to $I$.
After $v_1$ is moved to $I$, vertex $a_1$ will be moved to $M$ and the neighbors of $u$ together with $u$ will form a tail.
By reducing this tail by \refr{rule_tails}, at least two move $U$-vertices $v_2$ and $v_3$ will be removed from the graph,
which remains connected and contains at least one $M_0$-vertex $a_1$.
Hence at least four $U$-vertices will be eliminated.
Then we get a recurrence relation 
\eqn{4-7}{C(n)\leq C(n-7)+C(n-4).}

\subsection{Step~5(d)}

In Step~5(d), the algorithm chooses  a 5-cycle $uv_2a_1a_2v_3$ passing through
an  $M_0$-vertex $u$ which has at least three degree-2  neighbors,
and branches on a degree-2 vertex $v_1$ in $N(u)\setminus\{v_2,v_3\}$.
Denote $N(v_1)=\{u,b\}$, where  $b\in N_2(u)$.
Note that no pair of vertices in $N(uv_1)$ are adjacent  since otherwise $v_1$ would be i-reducible.
Also vertex $b$ cannot be a degree-2 vertex since
otherwise there would be a chain containing
a 2-path $uv_1b$.
Then $b$ is adjacent to at least two vertices in $N_2(uv_1)$.
We consider two subcases.
See \reff{case3} for an illustration of the neighbors of edge $uv_1$.

\vspace{-0mm}\fig{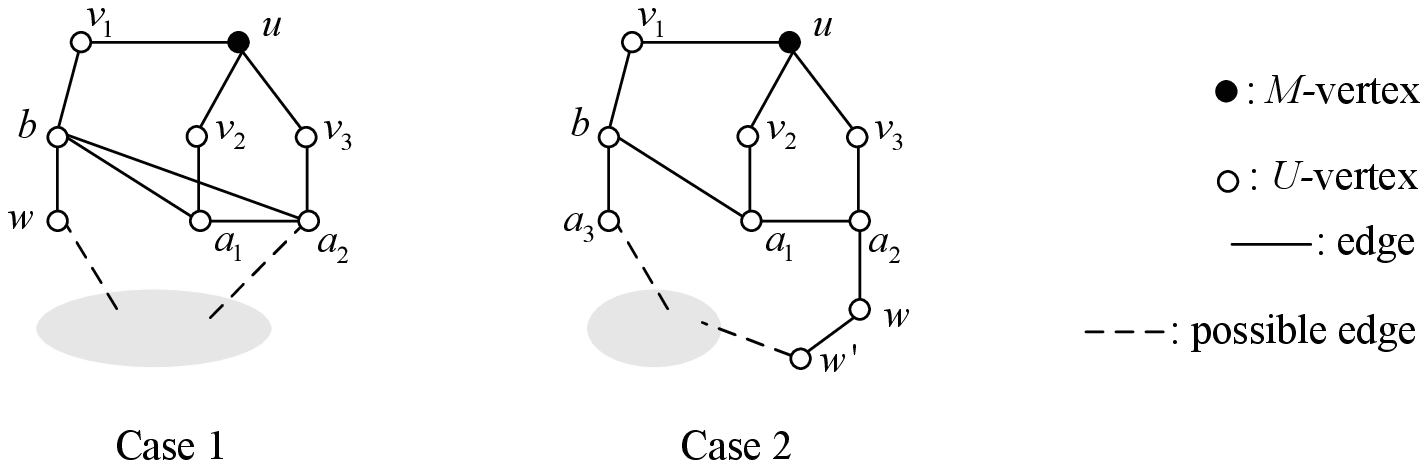}{0.9}{Neighbors of edge $uv_1$ in an instance at  Step~5(d)}{case3}\vspace{-0mm}

Case 1. $b$ is adjacent to both of $a_1$ and $a_2$:
Since the graph has more than six $U$-vertices,
we know that there is a vertex $w$ adjacent to some vertex in
$\{u,v_1,v_2,v_3,b,a_1,a_2\}$.
The vertex $w$ cannot be an $M$-vertex,
 because $N_2(v_1u)$  contains no $M$-vertices by \refl{withoutopt}(ii)
and if $w$ is an $M$-neighbor of $a_1$ or $a_2$ then $v_1$ would be i-reducible.
In the branch where $v_1$ is moved to $M$,
all the seven $U$-vertices $\{v_1,v_2,v_3,b,a_1,a_2, w\}$ will be eliminated.
In the other branch where $v_1$ is moved to $I$, vertex $b$ is moved to $M$.
Now \refr{rule_6cycle} can be applied to the 6-cycle $uv_2a_1ba_2v_3$.
Applying the reduction rule  eliminates at least two $U$-vertices $v_2$ and $v_3$.
This branch eliminates at least four $U$-vertices in $\{v_1,b,v_2,v_3\}$.
We get the same recurrence relation as \refe{4-7}.

Case 2. At least one of $a_1$ and $a_2$, say $a_2$ is not adjacent to $b$:
Vertex $b$ is adjacent to a vertex $a_3\in N_2(uv_1)\setminus \{a_1,a_2\}$.
Note that $a_2$ has no neighbor in $N(uv_1)\setminus \{v_3\}$,
since otherwise $uv_3a_2$ would be contained in a 4-cycle and $v_3$ would be i-reducible.
Also $a_2$ has no neighbor  in $N_2(uv_1)\setminus \{a_1\}$, since otherwise $v_1$ would be i-reducible.
It is also impossible that $a_2$ has no other neighbor than $a_1$ and $v_3$, since otherwise $v_2$ would be i-reducible.
Hence  $a_2$ has an $N_3(uv_1)$-neighbor, which must be
a $U$-vertex since otherwise $v_1$ would be i-reducible.
We know that either $a_2$ has two $U$-neighbors $w$ and $w'\in N_3(uv_1)$ or $a_2$ has only one $U$-neighbor $w\in N_3(uv_1)$.
For the latter case, $w$ should also have a $U$-neighbor $w'\not\in N(uv_1)\cup N_2(uv_2)\cup \{w\}$,
where $w'$ cannot be an $M$-vertex since otherwise $v_2$ would be i-reducible.

In the first branch where $v_1$ is moved to $M$, the $U$-vertices in the following set will be eliminated
$$\{v_1\}\cup N(uv_1) \cup N_2(uv_1)\cup \{w,w'\},$$
where $|N(uv_1)|\geq 3$ since $\{v_2,v_3,b\}\subseteq N(uv_1)$ and $|N_2(uv_1)\cap U|=|N_2(uv_1)|\geq 3$ since $\{a_1,a_2, a_3\}\subseteq N_2(uv_1)$.
In the second branch where $v_1$ is moved to $I$, vertex $b$ will be moved to $M$, and
two $U$-vertices $v_1$ and $b$ are  eliminated.
Therefore we get a recurrence relation
$$C(n)\leq C(n-9)+C(n-2).$$

We further look at the second branch of moving $v_1$ to $I$.
Let $G'$ be the graph after removing $v_1$ and moving $b$ to $M$.
Recall that any resulting $I$-vertices in $G'$  will be removed in Step 3.
In $G'$, vertices $u$ and $b$ are  $M_0$-vertices, where no tail is created in $G'$.
We distinguish two subcases.

(i) At least one of $u$ and $b$ is of degree 2 in $G'$:
Then the algorithm executes either
 some reduction operation other than reducing tails
to eliminate at least one more $U$-vertex
or   Step~5(a) or  (b) in $G'$
with a recurrence relation covered by \refe{2dbranch_g} in \refl{degree2}.
In the former, we obtain a recurrence relation
\eqn{9-3}{C(n)\leq C(n-9)+C(n-3).}
In the latter, we analyze a recurrence relation for the operation combined with
the branching on $v_1$ and branching in $G'$.
By \refl{combined_recurrences},
we can ignore recurrence relations covered by \refe{2dbranch_g},
and   we get a recurrence relation
\eqn{weak}{
\begin{array}{rcl}
C(n)&\leq& C(n-9)+C(n-2-8)+C(n-2-4)\\
&=&C(n-9)+C(n-10)+C(n-6).
\end{array}}

(ii) Neither of $u$ and $b$ is of degree 2 in $G'$:
We know that before branching on $v_1$,
vertex $u$ has at least four neighbors $v_1,v_2,v_3$ and $v_4$,
and vertex $b$ has at least two $U$-neighbors $a_3,a_4\in N_2(uv_1)$ such that $\{a_3,a_4\}\cap\{ a_1,a_2\}=\emptyset$.
 Therefore $|N(uv_1)|\geq 4$ and $|N_2(uv_1)\cap U|\geq 4$ in $G$.
In the first branch where $v_1$ is moved to $M$,
at least 11 $U$-vertices in $\{v_1\}\cup N(uv_1) \cup N_2(uv_1)\cup \{w,w'\}$ are eliminated.
We get a recurrence relation
\eqn{11-2}{C(n)\leq C(n-11)+C(n-2).}

\subsection{Step~5(e)}

In Step~5(e), the algorithm chooses
an  $M_0$-vertex $u$ and branches on a $U$-neighbor  $v_1$ of $u$,
where  $u$ is not contained in 5-cycles and $u$ has at least three neighbors $v_1, v_2$ and $v_3$
each of which is adjacent  to some vertex in $N_2(u)$.
Now all of $v_1,v_2$ and $v_3$ are degree-2 vertices.
For each $i=1,2,3$, let $a_i$ be the other neighbor of $v_i$ than $u$.
The six vertices in $\{v_1,v_2,v_3,a_1,a_2,a_3\}$ are different from each other,
because $u$ is not contained in a triangle or 4-cycle by \refl{reduced}.
Furthermore, vertex $a_1$ has at least two $N_2(v_1u)$-neighbors $b_1$ and $b_2$ since there is no chain, and
no neighbor of $a_1$ is in $\{v_2,v_3,a_2,a_3\}$ since $u$ is not contained in a 4-cycle or 5-cycle.
See \reff{case4} for an illustration of the neighbors of edge $uv_1$.
We distinguish two cases.

\vspace{-0mm}\fig{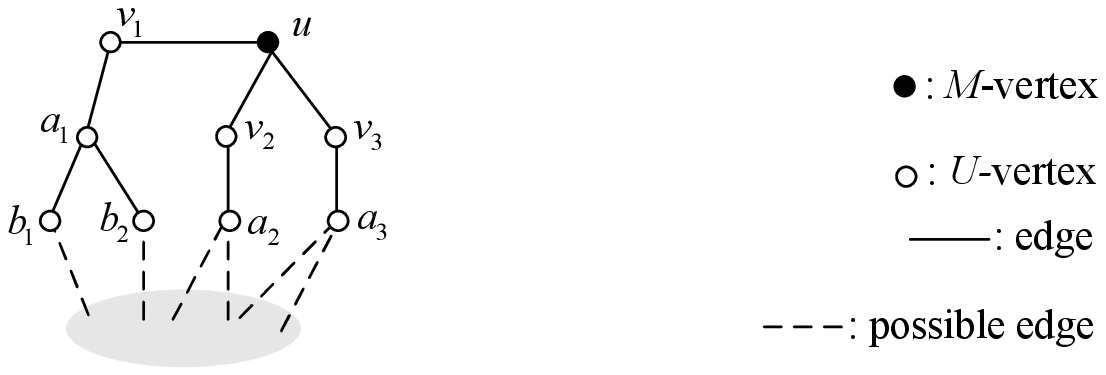}{0.7}{Neighbors of edge $uv_1$ in an instance at Step~5(e)}{case4}\vspace{-0mm}

Case~1. $u$ is of degree 3: We further distinguish two subcases.

(i)  $a_1$ has at least three $N_2(v_1u)$-neighbors $b_1, b_2$ and $b_3$:
Then
$N(v_1u)=\{a_1,v_2,v_3\}$  and $|N_2(v_1u)|\geq 5$ by $\{a_2,a_3, b_1,b_2,b_3\}\subseteq N_2(v_1u)$.
In the  first branch where $v_1$ is moved to $M$, $U$-vertices in $\{v_1\}\cup N(v_1u)\cup N_2(v_1u)$ will be eliminated,
 and the number of $U$-vertices
decreases by at least $1+3+5=9$.
In the second branch where $v_1$ is moved to $I$,  $a_1$ is moved to $M$,  and the number of $U$-vertices
decreases by 2, leaving a degree-$2$ $M_0$-vertex $u$ in the graph.
Analogously with Case~2 in Step~5(d), we can get either \refe{9-3} or \refe{weak}.

(ii)    $a_1$ has only two neighbors $b_1$ and $b_2\in N_2(v_1u)$:
Then  we can assume that  both of $a_2$ and $a_3$ are degree-3 vertices
by the choice of $v_1$ and $\max\{|N_2(a_2u)|,|N_2(a_3u)|\}\leq |N_2(a_1u)|$.
Now we have that  $|N(v_1u)|=3$ by $N(v_1u)=\{a_1,v_2,v_3\}$ and $|N_2(v_1u)|=4$ by $N_2(v_1u)=\{a_2,a_3, b_1,b_2\}$.
In the first branch of moving $v_1$ to $M$, at least eight $U$-vertices   will be eliminated.
In the second branch of moving $v_1$ to $I$,  at least two $U$-vertices   will be eliminated.
Then we have only a recurrence relation $C(n)\leq C(n-8)+C(n-2)$. 
We derive an improved recurrence relation
based on the fact that each of the generated instances has  a degree-2 $M$-vertex
but contains no tail:
In the first branch where $v_1$ is moved to $M$,  vertex
$a_2$ will be a degree-2 $M$-vertex; and  in the second branch where $v_1$ is moved to $I$,
vertex $a_1$ will be a degree-2 $M$-vertex.
It is easy to check that each of the instances still contains no tails.
To each instance, the algorithm  in the next step
 either eliminates at least one more $U$-vertex by reduction operations except reducing tails
or
branches with a recurrence relation covered by \refe{2dbranch_g} in \refl{degree2}.
When a  $U$-vertex is eliminated by a reduction rule in the first instance, analogously with the analysis in  Case~2 of Step~5(d),
 we get  recurrence relations
 \refe{9-3} and \refe{weak}.
 When the algorithm branches with a recurrence relation covered by \refe{2dbranch_g} in the first instance and eliminates one
 $U$-vertex by reduction operations in the second instance, we get
\eqn{weak2}{
\begin{array}{rcl}
C(n)&\leq& C(n-8-8)+C(n-8-4)+C(n-3)\\
&=&C(n-16)+C(n-12)+C(n-3).
\end{array}
 }
Otherwise the algorithm branches with a recurrence relation covered by \refe{2dbranch_g} in both of the two instances.
We show that the combined operation will create a recurrence relation covered by
\eqn{weakall}{
\begin{array}{rcl}
C(n)&\leq &C(n-8-8)+C(n-8-4)+C(n-2-8)+C(n-2-4)\\
&=&C(n-16)+C(n-12)+C(n-10)+C(n-6).                     
\end{array}
}

Assume that the algorithm branches with
 a recurrence $A:~C(n)\leq C(n-a_1)+C(n-a_2)$ and
a recurrence $B:~C(n)\leq C(n-b_1)+C(n-b_2)$ in the two instances, respectively,
where $A$ and $B$ are covered by \refe{2dbranch_g}.
By \refl{combined_recurrences}, we know that
recurrence $C_{AB}:~C(n)\leq C(n-8-a_1)+C(n-8-a_2)+C(n-2-b_1)+C(n-2-b_2)$ is covered by $C(n)\leq C(n-8-8)+C(n-8-4)+C(n-2-b_1)+C(n-2-b_2)$, and
recurrence
$C(n)\leq C(n-8-8)+C(n-8-4)+C(n-2-b_1)+C(n-2-b_2)$ is covered by $C(n)\leq C(n-8-8)+C(n-8-4)+C(n-2-8)+C(n-2-4)$.
Then $C_{AB}$ is covered by \refe{weakall}.

Case~2. $u$ is of degree $\geq 4$: Let $v_4$ denote the fourth neighbor of $u$,
 and $a_4$ denote the second neighbor of $v_4$.
Now we have that $|N(v_1u)|\geq 4$ by $\{a_1,v_2,v_3,v_4\}\subseteq N(v_1u)$ and $|N_2(v_1u)|\geq 5$
by $\{a_2,a_3,a_4,b_1,b_2\}\subseteq N_2(v_1u)$.
We distinguish two subcases.

(i) $a_1$ has only two $N_2(v_1u)$-neighbors $b_1$ and $b_2$:
Then by branching on $v_1$
we get a recurrence relation $C(n)\leq C(n-10)+C(n-2)$.
The branch of moving $v_1$ to $I$ leaves a graph with a degree-2 $U$-vertex $a_1$ and no tail.
Analogously with Case~2 in Step~5(d),
we can get two recurrence relations covered by \refe{9-3} and \refe{weak}, respectively.

(ii) $a_1$ has at least three neighbors $b_1,b_2$ and $b_3\in N_2(v_1u)$:
Then $|N_2(v_1u)|\geq 6$ by $\{a_2,a_3,a_4,b_1,b_2,b_3\}\subseteq N_2(v_1u)$.
In the branch where $v_1$ is moved to $M$, at least 11 $U$-vertices will be eliminated.
We can branch with \refe{11-2} in this case.

\subsection{Step~5(f)}
In Step~5(f), $M_0=\emptyset$ holds, and
the algorithm branches on a $U$-vertex $v_1$ of maximum degree,
where the degree of $v_1$ is  at least 3 since after Step~1 the maximum degree of the graph is at least 3.
Also in this step, the graph has no degree-1 vertex, otherwise the unique neighbor of each degree-1 vertex would be in $M_0$.
In the first  branch of moving $v_1$ to $M$, at least one $U$-vertex will be eliminated.
In the second branch of moving $v_1$ to $I$, all neighbors of $v_1$ will be moved to $M$ by applying reduction rules, and
the number of eliminated $U$-vertices is at least  $1+|N(v_1)|\geq 4$.
This leads to a recurrence relation $C(n)\leq C(n-1)+C(n-4)$.
To derive a better recurrence relation, we distinguish four cases.

Case 1. $v_1$ is contained in a triangle $v_1aa'$ or a 4-cycle $v_1abc$:
We show that the first  branch eliminates at least three $U$-vertices.
First consider the case where $v_1$ is contained in a triangle $v_1aa'$.
Let $a''$ be an $N(v_1)\setminus\{a,a'\}$-neighbor of $v_1$.
In the first branch of moving $v_1$ to $M$,
 all vertices in $N(v_1)\setminus\{a,a'\}$ will become i-reducible,
at least three $U$-vertices in $N[a'']$ will be eliminated.
Next consider the case where  $v_1$ is contained in a 4-cycle $v_1abc$.
In the first branch of moving $v_1$ to $M$, the two neighbors $a$ and $c$ of $v_1$ will become i-reducible,
and at least three $U$-vertices will be eliminated.
In any case, we can get a recurrence relation
$$C(n)\leq C(n-3)+C(n-4).$$

In what follows, we assume that $v_1$ is not contained any triangle or 4-cycle.

Case 2. $v_1$ has a degree-2 neighbor $a$:
Let $a'$ be the other neighbor of $a$, where $a'\not\in N(v_1)$
 since the condition of Case 1 does not hold.
Since the graph has no degree-1 vertex and $v_1$ is not contained in a 4-cycle,
we know that $a'$ has an $N(v_1)$-neighbor $w$.
In the second branch where $v_1$ is moved to $I$,
all vertices in $N[v_1]\cup \{a',w\}$ will be moved from $U$.
At least six $U$-vertices will be eliminated in this branch. Therefore we can get
$$C(n)\leq C(n-1)+C(n-6).$$

Case 3. $v_1$ is a vertex of degree $\geq 4$ with no degree-2 neighbors:
Since $|N(v_1)|\geq4$, it is easy to see that the algorithm  branches on $v_1$ with a recurrence relation
 $C(n)\leq C(n-1)+C(n-5)$.
We further look at the first branch, and denote $G'$ be the graph obtained by moving $v_1$ to $M$.
If at least one more $U$-vertex is moved to $M\cup I$ by a reduction operation applied to $G'$, then
we have a recurrence relation
$$C(n)\leq C(n-2)+C(n-5).$$
Assume that no more $U$-vertex is moved to $M\cup I$ in $G'$.
In $G'$,
vertex $v_1$ becomes an $M_0$-vertex with all neighbors of degree $\geq 3$,
and it is
an $M_0$-vertex such that all the neighbors are effective vertices.
In the next step, the algorithm will branch on an effective vertex adjacent to $v_1$ in $G'$
 with the recurrence relation  \refe{multi-eff}  with $d\geq 4$, i.e., $C(n)\leq C(n-12)+C(n-3)$,
by the analysis in Case~S2 in Step~5(a).
By combing this with the above recurrence relation, we get
$$C(n)\leq C(n-13)+C(n-4)+C(n-5).$$

Case 4. $v_1$ is a degree-3  vertex with no degree-2 neighbors:
By the choice of $v_1$, the current graph $G$  is a 3-regular graph without any triangle or 4-cycle.
Since $|N(v_1)|=3$,   the algorithm  branches on $v_1$ with a recurrence relation
 $C(n)\leq C(n-1)+C(n-4)$.
We further look at both branches.
Let $G_1$ and $G_2$ be the graphs obtained by moving $v_1$ to $M$ and $I$, respectively.

In $G_1$, all neighbors of $v_1$ will become effective vertices.
If an effective vertex in $G_1$ is  eliminated by a reduction rule, before the instance becomes a reduced one,
 then at least one more $U$-vertex will be moved to $M\cup I$ by the reduction operation.
On the other hand, the algorithm will branch on an effective vertex adjacent to $v_1$ in $G_1$
with the recurrence relation
  \refe{2dbranch} with $\lambda=1$, i.e., $C(n)\leq C(n-3)+C(n-9)$, by the analysis in Step~5(a).
In $G_2$,  all neighbors of $v_1$ are contained in $M$.
Each neighbor $v'$ of $v_1$ will become a degree-2 $M_0$-vertex in $G_2$ satisfying the condition of
Case~S1 in Step~5(a).
If no  more $U$-vertex is eliminated by reduction rules before the next branching,
then the algorithm will branch
with \refe{6-7} in $G_2$, where  \refr{rule_edge1}
cannot be applied to $G_2$ since the graph has no triangle.
If only one $U$-vertex is eliminated by reduction rules before
 the next branching, then the resulting graph still has at least one degree-2 $M_0$-vertex and the
algorithm will branch with \refe{2dbranch_g}.
Otherwise, two $U$-vertices are eliminated and the resulting graph still has an $M_0$-vertex.
Therefore we obtain six recurrence relations
\begin{eqnarray*}
\begin{array}{rcl}
C(n) &\leq & C(n-1-1)+C(n-4-6)+C(n-4-7)\\
&= & C(n-2)+C(n-10)+C(n-11),\\
& & \\
C(n)&\leq &C(n-1-1)+C(n-4-1-4)+C(n-4-1-8)\\
&=& C(n-2)+C(n-9)+C(n-13), \\
& & \\
C(n)&\leq &C(n-1-1)+C(n-4-2)=C(n-2)+C(n-6),\\
& & \\
C(n)&\leq &C(n-1-3)+C(n-1-9)+C(n-4-6)+C(n-4-7)\\
&=& C(n-4)+C(n-10)+C(n-10)+C(n-11),\\
& & \\
C(n)&\leq &C(n-1-3)+C(n-1-9)+C(n-4-1-4)+C(n-4-1-8)\\
&=& C(n-4)+C(n-10)+C(n-9)+C(n-13), ~~\mbox{and} \\
& & \\
C(n)&\leq &C(n-1-3)+C(n-1-9)+C(n-4-2)\\
&=& C(n-4)+C(n-10)+C(n-6).
\end{array}
\end{eqnarray*}


We have analyzed recurrence relations for all cases in Step~5.
In fact, the  recurrence relations in Step~5(f)  are not good enough to get our claimed running time bound.
However, the condition of Step~5(f) will not always happen.
By \refl{Mconnected} and \refc{cut}, we know that before creating a connected component $H$ of $G[M_0\cup U]$ that contains no $M_0$-vertex,
at least one tail adjacent to $H$ must have been removed from it
 except for the case where $H$ is the initial connected graph.
A tail contains at least two $U$-vertices.
Therefore, we see that
at least two more $U$-vertices  have been eliminated before Step~5(f) is executed.
Considering this, we can replace the above ten recurrence relations with
 \eqn{5-6}{C(n)\leq C(n-5)+C(n-6),}
 \eqn{3-8}{C(n)\leq C(n-3)+C(n-8),} 
  \eqn{4-7-2}{C(n)\leq C(n-4)+C(n-7),}
  \eqn{6-7-15}{C(n)\leq C(n-6)+C(n-7)+C(n-15),} 
  \eqn{4-12-13}{C(n)\leq C(n-4)+C(n-12)+C(n-13), } 
  \eqn{4-11-15}{C(n)\leq C(n-4)+C(n-11)+C(n-15),} 
    \eqn{4-8_1}{C(n)\leq C(n-4)+C(n-8),}
\eqn{6-12-12-13}{C(n)\leq C(n-6)+2C(n-12)+C(n-13),} 
 \eqn{6-11-12-15}{C(n)\leq C(n-6)+C(n-11)+C(n-12)+C(n-15),~~\mbox{and}} 
        \eqn{6-8-12}{C(n)\leq C(n-6)+C(n-8)+C(n-12).} 


\subsection{The final solution}

Among the above recurrence relations, the worst one with the largest branching factor is \refe{weakall}, which solves to $C(n)=O(1.1467^n)$. Then we get
\thmm{final}{\textsc{Dominating Induced Matching} can be solved in $1.1467^nn^{O(1)}$ time.}

\section{Concluding Remarks}\label{sec_con}
By designing several branching rules, we can eliminate $M_0$-vertices contained in some cycles of length at most 6.
This improves recurrence relations in the several previously worst cases in ~\cite{LMS:DIMexact}.
Finally, the worst case in our algorithm will be to branch on a vertex in a local graph
without any special structure and then we get the claimed running time bound.

In our algorithm, we use the number of undecided vertices as the measure.
For most NP-hard graph problems, the best exact algorithms are designed and analyzed by using the measure and conquer method,
which requires a complicated measure.
If we also introduce the measure and conquer method to our algorithm, we may need to set different weights to $U$-vertices.
However, we have not found any good weight setting scheme to improve the bound of the running time.
It leaves as a question whether or not the measure and conquer method is also helpful in designing exact algorithms for \textsc{Dominating Induced Matching}.


\textsc{Dominating Induced Matching} is to partition a graph into two parts $A$ and $B$ which
 induce an independent set and a matching, respectively.
For further study, we may consider algorithms for the extended problem:  for
integers $a,b\geq 0$, we are asked
to partition a given graph into two parts $A$ and $B$   that $A$
   induce a degree-$a$ graph and a degree-$b$ graph, respectively.
Some complexity results of this kind of extended problems can be found in~\cite{xn:BDG}.


\section*{Acknowledgement}
The first author was supported in part by National Natural Science Foundation of China under the Grant
61370071.


\begin{thebibliography}{99}
%
\bibitem{BHN:EEDhole-free}
Brandst\"{a}dt, A., Hundt, C., and Nevries, R.:
Efficient edge domination on hole-free
graphs in polynomial time. In:  LATIN 2010, LNCS 6034,  650--661, 2010

\bibitem{BM:DIMp7}
Brandst\"{a}dt, A., and Mosca, R.:
Dominating induced matchings for $P_7$-free graphs in linear time.
\emph{Algorithmica} (2013) DOI 10.1007/s00453-012-9709-4

\bibitem{BLR:EEDhypergraph}
Brandst\"{a}dt, A., Leitert, A., and Rautenbach, D.:
Efficient dominating and edge dominating sets for graphs and hypergraphs.
In: ISAAC 2012, LNCS 7676,  267--277, 2012.

\bibitem{CCDS:EEDRegularGraph}
Cardoso, D.M., Cerdeira, J.O., Delorme, C., and Silva, P.C.:
Efficient edge domination in regular graphs.
\emph{Discrete Applied Math.} \textbf{156} (2008), 3060--3065

\bibitem{CKL:DIM}
Cardoso, D.M., Korpelainen, N., and Lozin, V.V.:
On the complexity of the dominating induced matching problem in hereditary classes of graphs.
\emph{Discrete Applied Math.} \textbf{159} (2011), 521--531

\bibitem{Fomin:book}
Fomin, F. V.  and Kratsch, D.:
\newblock Exact Exponential Algorithms, Springer (2010)

\bibitem{GSSH:EED}
Grinstead, D.L., Slater, P.L., Sherwani, N.A., and Holmes, N.D.:
Efficient edge domination problems in graphs.
\emph{Information Processing Letters} \textbf{48}(1993), 221--228


\bibitem{LMS:DIMexact}
Lin, M.C., Mizrahi, M.J., and Szwarcfiter, J.L.:
An $O^*(1.1939^n)$ time algorithm for minimum
weighted dominating induced matching.
In: ISAAC 2013, LNCS 8283,  558--567, 2013.

\bibitem{LMS:DIMexact_odd}
Lin, M.C., Mizrahi, M.J., and Szwarcfiter, J.L.:
Exact algorithms for dominating induced matchings. CoRR, abs/1301.7602 (2013)


\bibitem{PandEED}
Lu, C.L., Ko, M.-T., and Tang, C.Y.:
Perfect edge domination and efficient edge domination
in graphs.
\emph{Discrete Applied Math.} \textbf{119} (2002), 227--250

\bibitem{EED:bpermutation}
Lu, C.L. and Tang, C.Y.:
Solving the weighted efficient edge domination problem on bipartite permutation graphs.
\emph{Discrete Applied Math.} \textbf{87}(1998), 203--211

\bibitem{kn:rooij}
Van Rooij,~J.~M. and  Bodlaender,~H.~L.
Exact algorithms for edge domination,
\emph{Algorithmica} \textbf{64}(4) (2012),  535--563

\bibitem{xn:eds}
Xiao,~M.  and Nagamochi, H.:
\newblock A refined exact algorithm for edge dominating set.
\newblock \emph{Theoretical Computer Science }(2014) DOI: 10.1016/j.tcs.2014.07.019

\bibitem{xn:BDG}
Xiao,~M.  and Nagamochi, H.:
\newblock Complexity and kernels for bipartition into degree-bounded induced graphs.
\newblock Manuscript. 2014 (Submitted)




\end{thebibliography}
\end{document}